\documentstyle[aaspp]{article}

\newcommand{\bq}{\begin{equation}}
\newcommand{\eq}{\end{equation}}

\begin{document}
\def\refitem{\par\parskip 0pt\noindent\hangindent 20pt}
 
\title{The Case for an Accelerating Universe from Supernovae \\ {\it To Appear as an Invited Review for PASP}}

$\ \ \ \ \ \ \ \ \ \ \ \ \ \ \ \ \ \ \ \ \ \ \ \ \ \ \ \ \ \
\ \ \ \ \ \ \ \ \ \ \ \ \ \ \ \ \ \ \ \ \ \
\ \ $Adam G. Riess\footnote{Space Telescope Science Institute, 3700 San Martin Drive, Baltimore, MD 21218}

\begin{abstract}

The unexpected faintness of high-redshift Type Ia supernovae (SNe Ia), as
measured by two teams, has been interpreted as evidence
that the expansion of the Universe is accelerating.  We review the
current challenges to this interpretation and
seek to answer whether the cosmological implications are compelling.
We discuss future observations of SNe Ia which could offer extraordinary
evidence to test acceleration.

\end{abstract}
subject headings:  supernovae: general$-$cosmology: observations

\vfill
\eject

\section{Introduction}

    Two teams have presented observational evidence from high-redshift
    type Ia supernovae (SNe Ia) that the expansion of the Universe is
    accelerating, propelled by vacuum energy (Riess et al. 1998;
    Perlmutter et al. 1999).  The primary evidence for this hypothesis is the 
faintness of distant SNe Ia relative to their expected brightness
    in a decelerating Universe.  The question we propose to answer in this
    review is whether the observations of 
distant supernovae {\it compel} us to conclude that
    the expansion is accelerating. 

\section{Supernova Measurements}

\subsection{Past Work}

   Supernovae (SNe) have a long history of employment in the quest to
   measure Hubble's constant (see Branch 1998 for review) and
   currently provide a column of support for a strong 
consensus that $H_0$=60 to 70
   km s$^{-1}$ Mpc$^{-1}$.  The history of utilizing supernovae to
   measure the time evolution of the expansion rate is far briefer.
   All initial proposals for using high-redshift SNe to constrain global
   deceleration recognized the necessity of an optical space telescope
   to collect the data.  Wagoner (1977) envisioned application of
   Baade's method or the the Expanding Photosphere method (Kirshner \& Kwan 1974) to measure the angular diameter distance of Type I
   (hydrogen-deficient) and Type II (hydrogen-rich) 
supernovae at $z=0.3$.   Colgate (1979) demonstrated even
   greater prescience, suggesting that SNe I at $z=1$ could
   be used as standard candles for ``determining the cosmological
   constant with greater accuracy than other standard candles.''
   Further thoughts by Tammann (1979) included the necessity to
   account for the redshift of the observed light using
   spectrophotometry of SNe in the ultraviolet (i.e., $K$-corrections), host
   galaxy extinction, and time dilation of the light curves. 

  Yet even
   before the launch of the {\it Hubble Space Telescope} ({\it HST}),
 a persistent
   two-year ground-based effort by a Danish group was rewarded by the discovery
   of their first (and only reported) high-redshift SN Ia,
   SN 1988U at $z=0.31$ (Norgarrd-Nielsen et al. 1989) as well as modest bounds on the deceleration parameter, $-0.6 < q_0 < 2.5$.  This team
   employed modern image processing techniques to scale the brightness
   and resolution of images of high-redshift clusters to match
   previous images and looked for supernovae in the difference frames.
   Unfortunately, the project's low discovery rate coupled with the
  dispersion of SNe Ia when 
treated as perfect standard candles ($\sim$0.5 mag) suggested
   that the determination of the deceleration parameter would require
   a scientific lifetime.  

   However, great progress was made by the Supernova Cosmology Project
   (SCP) in the detection rate of high-redshift 
 SNe Ia by employing large-format CCDs,
   large-aperture telescopes, and more sophisticated image-analysis
   techniques (Perlmutter et al. 1995).  These advances led to the
   detection of seven SNe Ia at $z\approx0.4$ between 1992 and 1994,
   yielding a confidence region that suggested
of a flat, $\Lambda=0$ universe but with a large range of uncertainty
   (Perlmutter et al. 1997).  

   The High-$z$ Supernova Search Team (HZT) joined the hunt for
   high-redshift SNe Ia with their discovery of SN 1995K at $z=0.48$
   (Schmidt et al. 1998).   Both teams made rapid improvements in
   their ability to discover ever greater numbers of SNe Ia at still
   larger
 redshifts.  One-time redshift record holders included SN
   1997ap at $z=0.83$ (SCP; Perlmutter et al. 1998), SN 1997ck at $z=0.97$
   (HZT; Garnavich et al. 1998a), SN 1998eq at $z=1.20$ (SCP; Aldering et
   al. 1998) and SN 1999fv at z=1.23 (HZT; Tonry et al. 1999). 
   Before either teams' claimed 
samples of high-redshift SNe Ia were large enough
   to detect the acceleration signal, both teams found
 the data to be inconsistent with a Universe closed by
   matter (Garnavich et al. 1998a; Perlmutter et al. 1998).  

\begin{table}[h]
\begin{center}
\vspace{0.4cm}
\begin{tabular}{llll}
\multicolumn{4}{c}{  Record redshift Type Ia Supernovae } \\
\hline
\hline
SN Ia & $z$ & discoverers & discovered \\
\hline
SN 1988U &      0.31 &  N\o rgaard-Nielsen-Nielsen et al;  IAUC 4641 & 8/1988 \\
SN 1992bi &    0.46 &    SCP; IAUC 5652 & 4/1992 \\
SN 1995K &	0.48 &   HZT; IAUC 6160 & 3/1995 \\
SN 1995at &     0.66 &   SCP; IAUC 6270 &   12/1995 \\
SN 1996cl &     0.83 &   SCP; IAUC 6621 & 3/1996  \\
SN 1997ap &	0.83 &   SCP; IAUC 6596 & 3/1997 \\
SN 1997ck$^*$ &	0.97 &   HZT; IAUC 6646 & 4/1997 \\
SN 1998eq &	1.20	& SCP; IAUC 7046 & 10/1998 \\
SN 1999fv &	1.23 &   HZT; IAUC 7312 & 11/1999 \\
\hline
\hline
\multicolumn{4}{l}{$^*$ Poor spectrum of SN but light curve
consistent with SN Ia at measured $z$ of host} \\
\end{tabular}
\end{center}
\end{table}

   These observational feats were preceded by increased understanding
   and ability to make use of SNe Ia observations to constrain the
   cosmological parameters.  Empirical correlations between SN Ia
   light curve shapes and peak luminosity improved the
   precision of distance estimates beyond the standard candle model
   (Phillips 1993; Hamuy et al. 1995; Riess, Press, Kirshner 1995;
   Perlmutter et al. 1995;
   Tripp 1997; Saha et al. 1999; Parodi et al. 2000).
  Studies of SN Ia colors provided
   the means to distinguish supernovae which were reddened by dust
from those which were intrinsically red (Riess, Press, Kirshner 1996;
   Riess et al. 1998; Phillips et al. 1998; Tripp \& Branch 1999).
   Goobar \& Perlmutter (1995) showed that
   measurements of SNe Ia at different redshift intervals could break
   degeneracies between $\Omega_M$ and $\Omega_\Lambda$.  Additional
   work on cross-filter $K$-corrections provided the ability to
   accurately transform the observations of high-redshift SNe Ia to
   the rest-frame (Kim, Goobar, \& Perlmutter 1996).

\subsection{Observations}

   In order to resolve whether the results from high-redshift SNe Ia
   are compelling it is important to review how the measurements were
   obtained.  

   Both the SCP and the HZT detected their samples of
   high-redshift SNe Ia by using large-format CCDs at telescopes with
   large apertures (most commonly the CTIO 4-m Blanco telescope).
   Using two sets of deep images in the $R$ or $I$-band spaced across
   a lunation, the ``template'' images are subtracted
   from the second-epoch images 
and automated software searches for sources in the
   difference images whose intensity surpasses a specific threshold.
   The observations taken in pairs are spaced over a small time
   interval of a few minutes to eliminate moving transients.
   Human inspectors ``filter'' the automated results in an
   effort to maximize the likelihood that candidates are supernovae (Schmidt et al. 1998; Perlmutter et al. 1999).
   Because the resources available for collecting spectral
   identifications for all the candidates are insufficient, candidates
   which appear most likely to be SNe Ia are given priority.  The
   factors which are favored in the human selection criteria include
   candidates which are separated from the host galaxy nuclei and those that
 show good contrast with the host galaxy (especially those with little or no
   apparent host).   The signal-to-noise ratios of the identifying spectra vary
   greatly (Perlmutter et al. 1995, 1998; Riess et al. 1998) but have
   improved with the availability of the Keck Telescope and
   an increased emphasis on the search for clues of SN Ia evolution.
   Most of the spectral identifications were made by visual comparison
   to  template spectra of nearby SNe Ia.  More recently automated
   cross-correlation techniques have been employed (Riess et
   al. 1997).  Approximately half of the
   SN Ia redshifts were determined from narrow emission lines or Ca H
   and K absorption; the rest were derived from the broad supernova
   features.  Although formal statistics are not currently available,
   we are aware that the SCP candidates have yielded a greater
   fraction of SNe Ia than the HZT candidates, a significant number
of which turn out to be SNe II.   Taken together, about half of the two
   teams' candidates are revealed to be SNe Ia, with the rest
   classified as SNe II, AGNs, flare stars, or unclassified objects.
   SNe Ia in the desired redshift range are monitored photometrically
   in two colors by observatories around the world using red-sensitive
   passbands; {\it HST} has monitored
   the light curves of about two dozen of these objects. 

     The initial search template images are eventually 
replaced by deeper
     template images obtained with good seeing and
    taken one or more years after discovery
     when the SNe Ia have faded by five to eight magnitudes.  More
     recently, repeated searching of the same fields has provided deep
     template images {\it before} the SN explosion.  After the SN images are
     bias corrected and flat-fielded they are geometrically aligned,
     and the resolution and intensity are scaled to match those of the
     templates.  After subtracting the host galaxy light the SN
     magnitudes are measured.  The SCP uses aperture photometry; the
     HZT fits point-spread-functions.  Uncertainties are determined
     synthetically by the injection of artificial SNe of known brightness.
  The SCP uses
     standard passbands and Landolt (1982) standards of comparison
     while the HZT uses a custom passband system which is transformed
     to the Landolt scale (Schmidt et al. 1998).  Custom cross-band
     $K$-corrections are calculated using spectrophotometry of nearby
     SNe Ia whose colors are reddened to match the high-redshift objects.

     The distances are measured by fitting empirical families of light
     curves to the flux observations of individual supernovae.  The measured distances are derived from
     the luminosity distance, \bq D_{L} = \left(\frac{{\cal L}}{4 \pi {\cal
F}}\right)^{\frac{1}{2}} \eq where ${\cal L}$ and ${\cal
F}$ are the SN's intrinsic
luminosity and observed flux, respectively and $D_L$ is in Mpc. 
  Alternately, a
     logarithmic luminosity distance (i.e., the distance modulus) is used:
\bq .     \mu=m-M=5\log D_L +25,\eq where $M$ is the SN's absolute
     magnitude and $m$ is the observed magnitude in a given passband.

  Three different light-curve fitting methods have been
     used to measure the distances, each of which determines the shape of
     the best-fitting light curve to identify the individual luminosity of the SN Ia.  The HZT has used both the multi-color light-curve shape method (MLCS; Riess, Press, \& Kirshner
     1996; Riess et al. 1998) and a template fitting approach based on
     the parameter $\Delta m_{15}(B)$ (Phillips 1993; Phillips et
     al. 1999), while the SCP uses the ``stretch method'' (Perlmutter
     et al. 1995; 1999). 

 Nearby SNe Ia provide both the measure of the Hubble flow and the
     means to calibrate the relationship between light-curve shape and
     luminosity.  The SCP uses $\sim$20 nearby SNe Ia in the Hubble
     flow from the Cal\'{a}n/Tololo Survey
     (Hamuy et al. 1996) 
     while the HZT adds to this set an equal number of SNe from the
     CfA Sample (Riess et al. 1999). 

     Dust extinction is handled somewhat differently by the two
     teams.  The HZT measures the extinction from the
     $B-V$ reddening (i.e., the color excess) and then combines this
     measurement in a Bayesian formalism using a prior host galaxy
     extinction distribution calculated by Hatano, Branch, \& Deaton
     (1998).  This treatment assumes that extinction 
     makes supernovae appear dimmer (farther), never brighter
     (closer).
  The SCP uses a number of
     different approaches (including the HZT approach), but favors
     making no individual extinction corrections and discarding
     outliers.  Figure 1 shows a single Hubble diagram made 
with the data from both teams (Riess et al. 1998; Perlmutter et al. 1999).  

     The measured distances are then compared to those expected for
     their redshifts as a function of the cosmological parameters
     $\Omega_M$, $\Omega_\Lambda$, and $H_0$:
\bq
 D_L= c H_0^{-1}(1+z)\left | \Omega_k \right |^{-1/2}sinn\lbrace\left | \Omega_k \right |^{1/
2}
\int_0^zdz[(1+z)^2(1+\Omega_Mz)-z(2+z)\Omega_\Lambda ]^{-1/2}\rbrace,
\eq
where $\Omega_k=1-\Omega_M-\Omega_\Lambda$, and $sinn$ is $\sinh$ for
$\Omega_k \geq 0$ and $\sin$ for $\Omega_k \leq 0$
(Carroll, Press, \& Turner 1992).  The likelihoods for cosmological
parameters are determined by minimizing the $\chi^2$ statistic
between the measured and predicted distances (Riess et al. 1998;
Perlmutter et al. 1999).   Determination of  $\Omega_M$ and
$\Omega_\Lambda$ are independent of the value of $H_0$ or the absolute
magnitude calibration of SNe Ia.  The confidence intervals in the
$\Omega_M$-$\Omega_\Lambda$ plane determined by both teams are shown
in Figure 2.  In this plane, acceleration is defined by the region
     where the deceleration parameter, $q_0$, is negative, \bq
q_0={\Omega_M \over 2} - \Omega_\Lambda < 0. \eq

\subsection{Results} 

     The visual impression from Figure 2 is that the supernova
     observations favor the parameter space containing a positive cosmological
     constant and an accelerating Universe with high statistical
     confidence.  The confidence regions determined by the two teams
     are in remarkably good agreement.  Both teams claim that these
     results do not result from chance with more than 99\% confidence
     (Riess et al. 1998; Perlmutter et al. 1999).  However, the
     specific confidence of these cosmological conclusions depends on
     which parameter space intervals one considers to be equally
     likely, {\it a priori}.  This point is addressed by Drell,
     Loredo, \& Wasserman (2000) who suggest that models with
     $\Omega_\Lambda$=0 and $\Omega_\Lambda \neq 0$ could be
     considered equally probable, {\it a priori}.
     While Riess et al. (1998) and Perlmutter
     et al. (1999) considered the probability that  $\Omega_\Lambda >
     0$ with a flat prior in a linear $\Omega_\Lambda$ space, an
     alternative is to use a flat prior in a {\it logarithmic}
     $\Omega_\Lambda$ space. 

     Another useful way to quantify the SN Ia constraints has been
     given by Perlmutter et al. (1999) as
     $0.8\Omega_M-0.6\Omega_\Lambda=-0.2\pm0.1$, a result which
     applies equally well to the Riess et al. (1998) data. 

     A more illuminating way to quantify the evidence for an
     accelerating Universe is to consider how the SN Ia distances
     depart from decelerating or ``coasting'' models.
  The average high-redshift SN Ia is 0.19 mag
     dimmer or $\sim$10\% farther than expected for a Universe with no
     cosmological constant and negligible matter ($\Omega_\Lambda=0$, $\Omega_M=0$).  Of course it's
     apparent that the Universe has more than negligible matter
     and the current consensus from the mass,
   light, X-ray emission, numbers, and motions of clusters of galaxies
is that $\Omega_M \approx 0.3$ (Carlberg et al. 1996; Bahcall, Fan, \& Cen 1997; Lin et al. 1996; Strauss \& Willick 1995).  The high-redshift SNe Ia
     from both teams are 0.28 mag dimmer or 14\% farther than expected
     in a Universe with this much matter and no cosmological constant.  
     The {\it statistical} uncertainty of these values is 0.08 and
     0.06 mag (or 4\% and 3\% in distance) for the HZT and SCP,
     respectively.  The observed dispersion of the high-redshift SNe Ia around the best-fit cosmology is 0.21 mag for the HZT and 0.36 mag for the SCP.  
A frequentist would consider the
     accelerating Universe to be statistically likely at the 3$\sigma$ to
     4$\sigma$ level (i.e., $>$ 99\%),
 while the Bayseian likelihood would depend
     on a statement of the natural space and scale for $\Omega_\Lambda$.

     Simply put, high-redshift SNe Ia are $\sim$0.25 mag 
    fainter than expected in our
     Universe with its presumed mass density but 
     without a cosmological constant (or which is not accelerating).
    The statistical confidence that SNe Ia are fainter than expected
    is high enough to accept
    that it does not result from chance and additional  
     SNe Ia continue to support this conclusion (Schmidt 2000).  Rather,
    this result is only challenged by systematic uncertainties not
    reflected in the variance of high-redshift SN Ia distance
    measurements.  In the
    following sections we review the challenges to the cosmological
    interpretation of the SN Ia observations and consider whether the evidence
    compels us to believe that the Universe is accelerating.

\section{Challenges and Tests of the Accelerating Universe}

    \subsection{Evolution}

       Could SNe Ia at $z=0.5$, a look-back time of $\sim$ 5 Gyr, be
       intrinsically fainter than nearby SNe Ia by 25\%? 
 For the purpose of using SNe Ia
       as distance indicators near and far, we are concerned only with
       an evolution which changes the luminosity of a SN Ia for a
       fixed light-curve shape.  Evolution is a major obstacle to the
       measurement of cosmological parameters,
       having plagued workers who tried to infer the global deceleration rate from brightest
cluster galaxies in the 1970s (Sandage \& Hardy 1973).    We will
       consider both the theoretical and empirical indications for SN
       Ia evolution.  

       Theoretical understanding of SNe Ia provides reasons
       to believe that evolution is not a challenge to the
       accelerating Universe.  SNe Ia are events which occur on stellar
scales, not galaxian scales, and therefore should be less 
subject to the known
evolution of stellar populations.
However, our inability to
       conclusively identify
 the progenitor systems (see Livio 2000 for a review)
       and our lack of a {\it complete} theoretical model (see
       Leibundgut 2000 for a review) means we cannot rely exclusively
       on theory to rule out the critical
       degree of evolution.

  Nevertheless, theoretical
       calculations can provide some insight into this question.  
       H\"{o}flich, Wheeler, \& Thielemann (1998) have calculated
models of spectra of SNe Ia with solar and one-third solar metallicities
and have found little difference between the spectral energy
       distribution over the wavelengths
 where the SNe Ia have been observed (see
       Figure 3).  In principle, changes in the age and hence initial
       mass of the progenitor star at high redshift could yield white
       dwarfs of varying carbon-to-oxygen (C/O) ratio.  It is currently
       difficult to assess if such a variation could produce
       significant evolution as these calculations lack the necessary
       precision (Dominguez et al. 1999; von Hippel, Bothun, \&
       Schommer 1997).  Umeda et al. (1999) suggest that a
       change in the C/O ratio is the source of the inhomogeneity in
       SN Ia luminosity, but they conclude 
that calibration of the luminosity via
       light-curve shape relations effectively inoculates the
       cosmological measurements to an evolution in the C/O ratio.

   To date, answers to the question of whether SNe Ia evolve have been
   sought from empirical evidence.  In the nearby sample, SNe Ia are
observed in a wide range of host-galaxy morphologies including
   ellipticals, post-starburst galaxies (e.g., SN 1972E in NGC 5253), low surface brightness galaxies (e.g., SN 1995ak in IC 1844), irregulars (e.g., SN 1937C in IC 4182), S0s (e.g., SN 1995D in NGC 2962), and early to late-type spirals (van den Bergh 1994; Cappellaro et al. 1997; Hamuy et al. 1996; Riess et
   al. 1999).  The range of metallicity, stellar age, and
interstellar environments probed by the nearby hosts is much greater than 
the mean evolution in these properties for individual 
galaxies between $z=0$ and
   $z=0.5$.    Some
   variation in the observed characteristics of SNe Ia with host
   morphology has been seen in the nearby sample (Hamuy et al. 1996;
   Branch, Romanishin, \& Baron 1996)   Yet {\it after
correction} for the light-curve shape/luminosity relationship and
extinction, the observed residuals from the Hubble flow do not
   correlate with host galaxy morphology (see Figure 4). 
In addition, Hubble flow residuals show no correlation with the projected
   distance from the host center (see \S 3.6).
 This evidence suggests that SN Ia distance estimates are 
insensitive to variations in the
supernova progenitor environment and is the strongest argument against
   significant evolution to $z=0.5$ (Schmidt et al. 1998).  However,
   this evidence is still circumstantial as we cannot be sure that the
   {\it local} environments of the SN Ia progenitors are similar to
   the {\it average}
   environments of the hosts.  Future studies which probe
   the local regions of nearby SNe Ia should be able to explore their variance.

  The other empirical test of evolution has been to compare the 
observed characteristics of low-redshift and
high-redshift SNe Ia.  The assumption of this test is that a
luminosity evolution of $\sim$25\% would be accompanied by other
visibly altered characteristics of the explosion.  Here too
our lack of firm theoretical footing makes it difficult to gauge the
correspondence between any evolution in distance-independent quantities
and luminosity.  Therefore we must conservatively demand that
observations of {\it all} observables of distant SNe Ia be
statistically consistent with the nearby sample. 

\subsubsection{Spectra} 

  Comparisons of high-quality spectra between nearby and
 high-redshift SN Ia, such as those seen in Figure 5, 
 have revealed remarkable similarity
(Riess et al. 1998; Perlmutter et al. 1998,
1999; Filippenko et al. 2000).  The spectral energy distribution is sensitive to the atmospheric conditions of the supernova (i.e.,
temperature, abundances, and ejecta velocities).  Even primitive modeling indicates that
 it would be
 difficult to
retain the primary features of the SN Ia spectrum while
altering the luminosity by about 20\% to 30\%.
Further, comparisons of temporal sequences of spectra reveal no apparent
differences as the photosphere recedes in mass (Filippenko et
 al. 2000), indicating that the superficial similarities persist at
 deeper layers.  However, among the variety of nearby SNe Ia
 are objects which are both $\sim$25\% fainter than the average and
 also display very typical spectral features (e.g., SN 1992A, see Figure
 5).  Therefore, the existing spectral test alone is not sufficient to check
 for this degree of evolution.

   While spectral similarity between nearby and distant SNe Ia
provides no indication of evolution, a {\it lack} of any spectral peculiarities among high-redshift SNe Ia could signal some changes at high redshift.   Li et al. (2000) find from the most unbiased survey of nearby SNe Ia to date that $\sim$ 20\% of SNe Ia are spectroscopically
similar to the overluminous SN 1991T.  SN 1991T-like objects show weak
Ca II, Si II, and S II, but prominent features of Fe III (Filippenko 1997), and close cousins, such as SN 1999aa, have similar characteristics with the exception that Ca II absorption is more normal.
Monte Carlo simulations of the
search criteria used by the SCP and the HZT team performed by Li,
Filippenko, \& Riess (2000) indicate that such overluminous objects
should comprise approximately 25\% of high-redshift SNe Ia (with some
uncertainty due to a possible link between such objects and
circumstellar dust).  To date, neither team has reported the existence
of a single SN 1991T-like object among $\sim$ 100 high-redshift
objects.  

 It is certain that the low signal-to-noise ratio of the
spectra of high-redshift SNe Ia, coupled with the redshifting of
spectral features out of the observable window makes it more difficult
to identify individuals from this peculiar class.  In addition, the spectroscopic peculiarities of SN 1991T-like objects are only apparent close to maximum light (or earlier), and some high-redshift SNe Ia may not have been observed early enough to identify their spectral peculiarities.  This same effect may also explain why the Cal\'{a}n/Tololo survey of 29 SNe Ia yielded no SN 1991T-like objects (Hamuy et al. 1996).  If, however, these observational biases are not to blame, the absence of SN 1991T-like SNe Ia at high redshift
could result from an evolution of the population of progenitor systems (see Livio 2000
for a review) or a subtle  difference in selection criteria
(see \S 3.6).  If true, this type of evolution may yield important
clues which help identify the progenitor systems (Ruiz-Lapuente \&
Canal 1998), but it is unlikely to affect the measurement of the
cosmological parameters since spectroscopically normal SNe Ia at low and high redshift have been used to derive the cosmological constraints.
 
\subsubsection{Broad-band}

  The distributions of light-curve shapes for nearby and distant SNe
Ia are statistically consistent (Riess et al. 1998;
Perlmutter et al. 1999).  Such consistency appears to extend to infrared light
curves of high-redshift SNe Ia which show the
characteristic second maximum of typical, low-redshift SNe Ia (Riess et
al. 2000).

An analysis by Drell et al. (2000) indicates that different light-curve
{\it fitting methods} may not be statistically consistent and that the
apparent differences may be a function of the light curve shape.
  However, these conclusions are highly sensitive to estimates of the
{\it correlated} distance uncertainties between 
different fitting methods and these correlated uncertainties are difficult to
estimate.  

Evolutionary changes in the model temperature and hence
the thermal output of the explosion could be detected from the colors
of pre-nebular supernovae.  The most significant analysis of 
$B-V$ colors, performed by Perlmutter et al. (1999),
 demonstrated consistency between
low and high-redshift SNe Ia at maximum light.  Likewise, Riess et al.
(2000) found that the $B-I$ colors of a SN Ia at $z=0.5$ were consistent
with those of nearby SNe Ia. 
 However, Falco et al. 1999 (see also McLeod et al. 1999) suggested that the
$B-V$ colors of high-redshift SNe Ia from the HZT may be
excessively blue, a conclusion which cannot be rejected by the $B-I$ color measurements by
Riess et al. (2000).  More data are needed to confirm or refute this
possibility.  If true, this could indicate either
evolution or the existence of a halo of Milky Way dust which would
redden the observed wavelengths of nearby SNe Ia more than redshifted
objects.  This latter possibility has been suggested by recent Milky Way dust maps of Schlegel, Finkbeiner, \& Davis (1999) in contrast to the previous maps of Burstein \& Heiles (1982), but it would augment rather than fully explain the faintness of distant SNe Ia.
 
  The risetime (i.e., the time interval between explosion and maximum light)
is sensitive to the ejecta opacity and the distribution of
$^{56}$Ni .  The risetime of nearby SNe Ia (Riess et
al. 1999b) and the high-redshift SNe Ia of the SCP 
(Goldhaber 1998; Groom 1998) were
initially strongly discrepant (Riess et al. 1999c).   However, a reanalysis
of the SCP high-redshift data by Aldering, Knop, \& Nugent (2000) finds the high-redshift risetime to be longer and much
more uncertain than indicated by Groom (1998).   The remaining difference
could be no more than a $\sim$2.0$\sigma$ chance occurrence.
More early photometry of distant SNe Ia is needed to increase the
significance of this test of evolution.

   Evolution is arguably the most serious challenge to the
cosmological interpretation of high-redshift SNe Ia.  Further studies, currently underway,
seek to compare the host galaxy morphologies and luminosity versus
light-curve shape relations for nearby and distant SNe Ia.  The results
reviewed in this section do not appear to provide any clear evidence of evolution.
  However, absence of evidence
is not necessarily evidence of absence.  The paucity of high
signal-to-noise ratio observations of high-redshift SNe Ia and the current lack of a
comprehensive theoretical model or a well-understood progenitor system keeps
the embers of skepticism aglow.  

\subsection{Dust}

   Consideration of a non-cosmological explanation for the
dimming of distant supernova light must invariably turn to a 
famous pitfall of optical astronomy: extinction.
Trouble has often followed dust in astronomy, a point first appreciated
   by Trumpler (1930) when analyzing the spatial distribution of
   Galactic stars.

\subsubsection{Ordinary Dust}

  An additional $\sim$25\% opacity of visual light by dust in the light paths of distant supernovae would
be sufficient to nullify the measurement of the accelerating Universe.  
  Both teams currently
measure SN Ia colors to correct for the ordinary kind of interstellar
extinction which reddens light.  Galactic extinction maps from
Burstein \& Heiles (1982) and Schlegel, Finkbeiner, \& Davis (1998)
were used by the HZT and the SCP, respectively, to correct individual SNe Ia 
for Milky Way extinction.  Such corrections were typically
less than 0.1 mag due to the high galactic latitudes of the SNe Ia.  Even a previously unknown halo of Galactic
dust would dim the restframe light of 
nearby SNe Ia more than highly redshifted SNe Ia and would therefore
not explain the cosmological indications.  

Measurements of $B-V$
colors have been used by both teams to test for and remove host galaxy
extinction (Riess et al. 1998; Perlmutter et al. 1999; see Figure 6).  Totani \& Kobayashi (1999) have suggested that the remaining uncertainty in the
mean measured $B-V$ color excess ($\sigma=0.02$ mag; Perlmutter et
al. 1999), 
when multipled by reddening ratios of
3 to 4 to determine the optical opacity, may be too large to discriminate between open and lambda-dominated
cosmologies with high confidence.  However, such concern seems
unwarranted as this uncertainty remains 3
to 4 times smaller than the size of the cosmological effect of an accelerating Universe.   

 Another measurable effect of the critical amount of mean interstellar
extinction is that it would introduce
 more dispersion in the distance measurements
than is currently observed.  A random line of sight into a host galaxy will
intersect a nonuniform amount of extinction.  Hatano, Branch, \&
Deaton (1998) have calculated the expected distribution of extinction
along random lines of sight into host galaxies.   A mean, uncorrected
extinction of 0.25 mag would induce twice the distance dispersion
observed by the HZT (Riess et al. 1998).  In addition, high-redshift
surveys are biased towards finding SNe Ia which have even less
extinction than would be expected from the distributions of Hatano et
al. (1998).

A more powerful way to search for reddening by dust is to
observe high-redshift SNe Ia over a large wavelength span: from the optical to the infrared.  Infrared color excesses would be more than twice as large as
$E_{B-V}$ for ordinary dust.  A set of such observations for SN 1999Q
($z=0.46$) disfavor $A_V=0.25$ mag of dust with Galactic-type
 reddening at high confidence (Riess et al. 2000), but more SNe Ia need
to be observed in the near-infrared to strengthen this conclusion.

\subsubsection{Grey Dust}

  More pernicious than ordinary dust is ``grey'' dust which could leave
  little or no imprint on the spectral energy distribution
  of SNe Ia.  Perfectly grey dust is only a theoretical construct, but
  dust which is greyer than Galactic-type dust (i.e., larger reddening
  ratios) does exist (Mathis 1990) and could
  challenge the cosmological interpretation of high-redshift SNe Ia.  

  Grey dust can be made with large spherical dust grains or elongated
  ``whiskers.''  Past studies of whiskers (Aguirre 1999a; Rana 1979, 1980)
  indicate that they would distort the cosmic microwave background (CMB),
 an effect which has
  not been seen. 
Like non-grey extinction, grey interstellar extinction does not provide an acceptable
explanation for the dimness of SNe because the inherent variations in the opacity along random lines of sight would induce more distance dispersion than is observed (Riess et al. 1998).
 
  Grey {\it intergalactic} extinction could affect measurements of the
  deceleration parameter (Eigenson 1949) without tell-tale dispersion
  or reddening.   Indeed, observations of neither SNe Ia nor other
astrophysical objects rule out a 30\% opacity by large semi-spherical
  dust grains (Aguirre 1999b).
Aguirre (1999a,b) has shown that a uniformly distributed component
of intergalactic grey dust with a mass density of
$\Omega_{dust}\approx 5\times10^{-5}$ and graphite grains of size
 $> 0.1 \ \mu$m could explain the faintness of high-$z$
SNe Ia without detectable reddening and without overproducing the
  currently unresolved portion of the
  far-infrared (far-IR) background.  However, this physical model of
  dust would provide some reddening which can readily be detected with observations in the optical and infrared.  Measurements by Riess et al. (2000; see Figure 7) of $E_{B-I}$ for a
  single high-redshift SN Ia disfavor a 30\% visual opcaity of grey dust at the $\sim 2.5 \sigma$
  confidence level, but more observations are needed to strengthen
  this conclusion.  Additional studies of the faint far-IR sources seen with
  SCUBA may soon provide definitive constraints on the unresolved component
  of the far-IR background and the viability of extragalactic grey
  dust.

\subsection{Gravitational Lensing}

   The inhomogeneous distribution of matter in the Universe typically
   deamplifies and very rarely amplifies the observed brightness of
   distant SNe Ia compared to the average.  (Note that the {\it mean} observed
   brightness must equal the
   unamplified value expected in a perfectly smooth Universe.)

   The size of the typical deamplification is a function
   of the SN Ia redshift, the mass density of the Universe and the fraction of dark matter locked into compact objects.
This effect has been quantified by a wide range of techniques (Kantowski, Vaughan, \& Branch 1995; Frieman 1997;
   Wambsganss et al. 1997; Holz \& Wald 1998; Kantowski 1998;
   Metcalf 1999; Barber 2000).   The effect of weak lensing on the observed distribution of luminosities of SNe Ia at $z=1$ and $z=0.5$ can be seen in Figure 8.  In the most relevant regime for the current
   SNe Ia at $z=0.5$ (i.e., $\Omega_M \approx 0.3$, and mostly diffuse
   dark matter) the typical deamplification is $\sim$2\%, much smaller
   than the cosmological effect.  An extreme case 
  (i.e., $\Omega_M \approx 0.5$, all matter in point masses) could
   deamplify the median SN Ia at $z=0.5$ by 5\% (Holz 1998), but this model is
   unlikely to be correct and the effect is still
 not large enough to negate the
   cosmological interpretation of high-redshift SNe Ia.  Perlmutter et al. (1999) considered lensing by up to $\Omega_M=0.25$ in compact material in the
   determination of their confidence intervals.  They found little
   impact on the likelihood of a positive cosmological constant (see
   Figure 9).   Wang (2000) suggests that by flux-averaging (i.e., 
binning the SNe Ia distances by redshift) one can reduce the bias due to
   weak lensing.  In the future, any bias due to weak lensing will naturally vanish as the sample sizes become larger and the mean observed luminosity more robust.  It is interesting and potentially useful to note
   that the observed distribution of SN Ia distances (see Figure 8) 
can in principle be used to
   determine the fraction of gravitating matter contained in
   compact objects (Seljak \& Holz 1999; Metcalf \& Silk 1999).

\subsection{Measurement Biases}

   In this section we consider if biases in the measurement process of
   high-redshift SNe Ia could mimic the evidence for an accelerating
   Universe.  An exhaustive list of such biases has been considered by
   Hogg (2000) and Hogg \& Turner (1998).  Here we discuss how these
   biases may apply to the supernova measurements.

   The observational challenge is to measure the distance to
   high-redshift SNe Ia which are 6 to 7 magnitudes fainter and have
   lower signal-to-noise ratio than those
   which delineate the Hubble flow.  Differences in the way low and high-redshift SNe Ia are observed must not introduce biases in their distance measures at more than the few percent level.  Indeed, Hogg (2000) has
   noted that the proximity of high-redshift SNe Ia to any
   reasonable world model is a testament to the feasibility of measuring
   distances across such a large range.  However, because the
   goal of these observations is precision cosmology (and not simply
   to demonstrate the dynamic range of useful photometry)
 our scrutiny must be greater.

   Charge transfer inefficiency (CTI) and detector non-linearities can cause
   faint objects to appear fainter.   However, ground-based
   observations of high-redshift SN Ia are limited by the bright sky, a regime
   in which these effects are widely found to be negligible.
   For space-based observatories such as the {\it Hubble Space Telescope},
   CTI is far more troublesome but quite correctable (Whitmore, Heyer, \&
   Casertano 1999).  In addition, only a subset of the
   high-redshift SNe Ia have been measured with {\it HST} and the
   cosmological conclusions do not depend on the inclusion of these objects.

   High precision, flux-conserving algorithms have been developed
   to properly subtract images of the host galaxy from images with SN
   light (Alard \& Lupton 1998).  Correctly employed, these methods reduce any
   biases in the measurement of the SN brightness to less than a few
   percent.  Tests for measurement biases and estimates of uncertainty
   are performed by both teams by 
the injection of artificial SNe into the observed
   images.

   Hogg \& Turner (1998) discuss a bias towards higher observed fluxes
   which naturally occurs when measuring the brightness at discovery of
   low signal-to-noise ratio sources.
  This bias results
   from the preferential selection of faint sources on the bright side
   of the Poisson distribution of photon statistics.  Follow-up
   observations of the source would not incur this bias.  This effect would
have little impact on the supernova distances measured by the HZT and SCP
   because the light curves are dominated by observations made {\it after}
   discovery.  In addition, the direction of this effect is opposite
   to the signal of an accelerating Universe.  However, this effect
   may become more important for SNe Ia found at $z > 1$ for which the discovery observation may provide one of the most significant measurements. 

\subsection{Selection Biases}

    Do the HZT and SCP preferentially select faint
SNe Ia at high redshift?
  Because we have already considered evolution in \S 3.1, here
we are only concerned with the characteristics of a high-redshift
sample which is drawn from the {\it same} population as the nearby
sample.  In so doing, we must also consider if the nearby sample is a
fair representation of that population.   

   As an example, consider the set of nearby SNe Ia which appear
   fainter than expected for their redshift in the bottom panel of
   Figure 1.  Presumably these objects appear dim due to
   the intrinsic random scatter of SNe Ia.  If, however, these SNe Ia
   had a characteristic in common which, in addition, 
 favored their discovery at
   high-redshift, a bias would result.  To date, no such characteristic has been identified and the observed dispersion of nearby SNe Ia is consistent with their measurement errors.    

   Howell, Wang, \& Wheeler (2000) found a difference between 
the projected distances from the hosts' centers
 for the nearby and distant SNe Ia (see Figure 10).  Many of the
   nearby SNe Ia were found in the photographic Cal\'{a}n/Tololo survey in
   which saturated galaxy cores masked
   SNe near their hosts' centers (Shaw 1979; Hamuy \& Pinto 1999).
   The result is that distant SNe
   Ia are more centrally located than the nearby sample.  However, in an
   analysis of 44 nearby SNe Ia,
   Riess et al. (1999) found no dependence of the distance measurement
   on the projected distance from the host center,
 so this selection effect appears to have
 no bearing on the cosmological use of SNe Ia (see Figure 11).

   Malmquist bias (Malmquist 1924; 1936) can shift the mean distance too close in a
   magnitude-limited survey of SNe Ia.  This effect seems
   to contrast with the cosmological dimming perceived in an accelerating
   Universe.  However, if the {\it nearby} sample were more afflicted by Malmquist bias than the
   distant sample, this bias could mimic an accelerating Universe.
 Because the intrinsic scatter of SN Ia distances is low
   ($\leq 0.15$ mag), Malmquist bias, which scales with the square of the
   dispersion (see Mihalas \& Binney 1981 for a derivation), is small
   for SNe Ia.  Perlmutter et al. (1999) made analytic calculations of
   Malmquist bias arising from the intrinsic dispersion of SNe Ia
   (assumed to be $0.17$
   mag) and the SCP search incompleteness (determined empirically) to
   estimate that the {\it net} bias between the samples is no more
   than 0.03 mag.  [Perlmutter et al. (1999) notes that the net bias
   may actually be closer to zero due to a compensating bias against the
   selection of light curves which are ``fast'' for their luminosity
   and therefore spend less time above the
   detection limit.]  Riess et al. (1998) used a Monte Carlo exercise
   to simulate the selection of SNe Ia near and far.  Inputs to this
   exercise included the time interval between successive search
   epochs, limiting magnitudes, observed light-curve shapes, and the
   distribution of SN Ia luminosities.  They
   report a net bias of less than $0.01$ mag.  These results indicate that the {\it net} Malmquist bias has negligible impact on
   the cosmological conclusions.   

\subsection{Alternative Cosmological Models}

   The conclusions drawn from high-redshift SNe Ia are predicated on a
   model with two free parameters, $\Omega_M$ and $\Omega_\Lambda$,
   and a Friedmann-Robertson-Walker cosmology.
   In the absence of a sound fundamental motivation for
   $\Omega_\Lambda \approx \Omega_M$, alternate and more general
   descriptions of an energy density with negative pressure have been
   suggested (Caldwell, Dave, \& Steinhardt 1998).  These
   phenomenological or ``quintessence'' models invoke a decaying
   scalar field rolling down a potential as the source of today's
   acceleration (Wang et al. 2000).  A distinction of these models
   from a cosmological constant is that $w$, the ratio of pressure to
   energy density, is between -1 and 0, whereas $w$ is exactly -1 for a
   cosmological constant.  For feasible quintessence models, $w$, the equation-of-state parameter, varies slowly with time
   and can be approximated today by a constant equal to
\bq \widetilde w \approx \int da \, \Omega_Q(a) \, w(a) / \int da \,
   \Omega_Q(a),
\eq where $a$ is the scale factor and $\Omega_Q$ is the energy density
of the vacuum component.  The current acceleration for these models
(assuming only two significant energy components today, $\Omega_M$ and
$\Omega_Q$) is
\bq q_0=   -\ddot a(t_0) a(t_0)/\dot a^2(t_0)={1\over
2}(\Omega_M+\Omega_Q(1+3w)), \eq and is generally less than for a
cosmological constant (all other parameters fixed).  Inspection of
equation (2) reveals that the Universe is accelerating if $q_0$ is
negative ($\ddot a(t_0)$ is positive), requiring that $w < -{1 \over 3}$,
independent of the value of $\Omega_M$.

   Can we determine if the expansion is accelerating in a quintessence
   model?   The SN Ia
   data from Perlmutter et al. (1999) and Riess et al. (1998) already
   provide meaningful constraints on $w$ (Garnavich et al. 1998b;
   Perlmutter et al. 1999).  Increasing the
   value of $w$ from -1 (i.e., for a cosmological constant) reduces
   the acceleration provided by a fixed value of $\Omega_Q$, but larger
   values of $\Omega_Q$ are needed to retain an acceptable fit
   to the data.  Graphically, increasing $w$ from -1 rotates
   the error ellipses in Figure 2 to favor lower values of $\Omega_M$
   and greater values of $\Omega_Q$. As seen from
   equation (6), the line separating an accelerating and decelerating
   Universe rotates in the same direction (always anchored at
   $\Omega_M=0$ and $\Omega_Q=0$), providing 
   no gain  
on an acceptable region of parameter space which is not
   accelerating.  The nearest intersection between a
   non-accelerating region of parameter space and one which is
   preferred by the data remains when $\Omega_M << 1$ and $\Omega_Q <<
   1$.  However, values near $\Omega_M=0$ and $\Omega_Q=0$ are poor fits to the
   data independent of the value of $w$. 

   Perhaps the simplest way to understand why the SN Ia data favor an
   accelerating Universe is to consider an FRW cosmology with
   $\Omega_M=0$ and no vacuum energy.
  This empty Universe must be neither accelerating nor
   decelerating but simply coasting.  The fact that the high-redshift
   SNe Ia are systematically farther for their redshift than expected
   in this cosmology means that the distance between low and
   high-redshift SNe Ia (where redshift is a surrogate for time) grew
   {\it faster} than expected for a Universe which has been coasting on
 today's Hubble expansion.  This implies that the Universe has
   been accelerating.

   An alternate cosmological explanation to acceleration has been
   posited by Goodwin et al. (1999) and Tomita (2000).  They suggest that the supernova
   data are also consistent with a decrease in the Hubble expansion by
   10-20\% beyond $z=0.1$ ($300 h^{-1}$ Mpc).  The distance at which
   the Hubble expansion dips
   would correspond to the approximate radius of the ``local''
   underdensity in which we live.  Although a few peculiar
   flow surveys support bulk motions on scales up to half this size
   (Lauer \& Postman 1994; Hudson et al. 1999), most recent surveys
   do not (Dale et al. 1999; Courteau et al. 2000; Colless et
   al. 1999; Riess et
   al. 1999;
   see Willick 2000 for a review).  However, the biggest problem with
   such a commodious, local underdensity is its great improbability.
   Power spectra demonstrate (Watkins \& Feldman 1995; Feldman \&
   Watkins 1998) 
that the density of the Universe is extremely
   homogeneous on this scale, and finding ourselves in the midst of
   such a vacuous location would be virtually anti-Copernican.  Using cold dark matter power spectra constrained
by CMB observations and large scale structure, Shi and Turner (1998) and Wang, Spergel, and Turner (1998) expect 0.5\% to
1.5\% variations in the Hubble constant on $300 h^{-1}$ Mpc scales, a factor of 20 times smaller than required in the local void model.
By filling in the Hubble diagram of SNe Ia at $0.1 < z < 0.2$ it would
   be possible to test this model.

   Outside the FRW cosmologies the SN Ia data can have significantly
different interpretations.  For example, in steady-state cosmologies,
SN redshifts do not come from expansion, but rather through
``tired-light'' processes.  However, the SN Ia data exhibit the
time dilation effect expected in an expanding Universe, implying that the tired-light hypothesis is incorrect (Leibundgut et al. 1996;
Goldhaber et al. 1996; Riess et al. 1997; but see Narlikar \& Arp 1997).  In the quasi-steady state cosmology,  the SN Ia data lead
to modifications of the model, such as matter creation during periodic expansion phases (Hoyle, Burbidge, \& Narlikar 2000).    A detailed consideration of how to interpret the SN Ia data in non-FRW cosmologies 
is beyond the scope of this
   review, but is thoroughly addressed by Hoyle, Burbidge, \& Narlikar (2000).
Alternative theories of gravity such as modified Newtonian dynamics models (MOND; Milgrom 1983, 1998; McGaugh \& de Blok 1998a,b) could also modify the interpretation of the observations of high-redshift SNe Ia.

\section{Conclusion}

   After reviewing the cosmological interpretation of SN Ia
   observations and the current challenges to the analysis of the
   data, we can now offer an answer to the question initially posed:  
   do the observations of 
distant supernovae {\it compel} us to conclude that
    the expansion of the Universe is accelerating?

  With full consideration of the evidence, we conclude that an
  accelerating Universe remains the most {\it likely} interpretation of the
  data because the alternatives, individually, appear less likely.  However,
  the quantity and quality of the SN Ia evidence {\it alone}
 is not yet sufficient to
  {\it compel} belief in an accelerating Universe.  The primary sources of
  reasonable doubt are evolution and extinction, as discussed above.
  Although the types of studies also described above could potentially yield
  evidence that either of these non-cosmological contaminants is
  significant, the current absence of such evidence does not suffice as
  definitive evidence of their absence.  Our current inability to identify the
  progenitors of SNe Ia and to formulate a self-consistent model of
  their explosions exacerbates such doubts.  Even optimists would
  acknowledge that neither of these theoretical challenges is likely to be met in the near future.

  Fortunately there are at least two routes to obtain compelling
  evidence to accept (or refute) the accelerating Universe, one of
  which employs the use of SNe Ia at even greater redshifts.  

\section{Epilogue}

\subsection {The Era of Deceleration}

   If the Universe is accelerating, it is a rather recent phenomenon
   likely commencing between $z \approx 0.4$ and 1.  Before this time the
   Universe was more compact and the pull of matter dominated the push of 
vacuum energy in the equation of motion.   As a result, the
   Universe at $z \geq 1$ must be {\it decelerating}.  This cosmological
   signature should
   be readily apparent by extending the Hubble diagram of SNe Ia
   to $z \approx 1.2$.  By this redshift SNe Ia in an accelerating Universe will
   cease to diverge in distance from an equally
   massive Universe without vacuum energy.   Alternatively, if a
   monotonically increasing, systematic effect is the source of the
   excessive faintness of high-redshift SNe Ia, the measured distances
   of SNe Ia at $z \geq 1$ will continue to diverge from a cosmology
   without vacuum energy and in addition would diverge from the
   cosmological model inferred from SNe Ia at $z=0.5$ (see Figure 12).
   Complex parameterizations of evolution or extinction selected to match
   both the accelerating and decelerating epochs of expansion would
   require a near conspiracy of fine tuning and are highly doubtful.
  
   Efforts are already underway to find and measure SNe Ia at $z > 1$.
   Gilliland, Nugent, \& Phillips (1999) used a subsequent epoch of
   the Hubble Deep Field to detect two SNe, one (SN 1997ff) 
with a photometric redshift of $z = 1.32$.  
   The elliptical host of SN 1997ff suggests that
   this object is of Type Ia, but the observations are insufficient to
   provide a useful distance estimate.  The SCP reported the discovery
   of SN 1998ef at $z=1.20$ (Aldering et al. 1998) and follow-up
   observations with the {\it HST} will provide a useful distance
   estimate (Aldering 2000).  The HZT recently
   reported the discovery of four SNe Ia at $z > 1$ including SN
   1999fv at $z=1.23$ (Tonry et al. 1999).
  From this growing sample will likely come the
   means to search for the epoch of deceleration.

 \subsection{Cosmic Complements}

   We previously sought to determine if the observations of SNe Ia {\it
   alone} require an accelerating Universe.  Now we will briefly
 consider the
   cosmological constraints provided by other astrophysical
   phenomena. A thorough discussion of these constraints is beyond the
   scope of this review but can be found elsewhere (Tyson \& Turner 1999; Roos \& Harun-or-Rashid 2000)

   Current measurements of the cosmic microwave background (CMB) power
   spectrum indicate that the sum total of energy densities is within
   10\% of
   unity.  This result is seen from the BOOMERANG 
   (Melchiorri et al. 1999; Lange et al. 2000; de Bernardis et al. 2000), the TACO (Miller et al. 1999) and MAXIMA experiments (Hanany et al. 2000; Balbi et al. 2000)
   and from a compilation of all other CMB measurements (Tegmark \& Zaldarriaga
   2000).  In addition, estimates of $\Omega_M$ from the mass,
   light, X-ray emission, numbers, and motions of clusters of galaxies
   converge around 0.2 to 0.3 (Carlberg et al. 1996; Bahcall, Fan, \& Cen 1997; Lin et al. 1996; Strauss \& Willick 1995).  These two pieces of information alone indicate a
   significant contribution by vacuum energy, sufficient to produce an
   accelerating Universe (see Figure 13).  Additional constraints from
   observations of the Lyman-alpha forest, cluster evolution, double
   radio galaxies, and statistics of gravitational lenses have been
   used to tighten these conclusions (Roos \& Harun-or-Rashid 2000;
   Turner 1999; Eisenstein, Hu, \& Tegmark 1999; Lineweaver 1998).  

   Although no single cosmological observation yields a conclusive
   census of the energy densities in the Universe, the combined
   constraints from multiple experiments is providing strong bounds
   on the cosmological parameters.  Each individual experiment has
   unique sources of systematic uncertainty.   By combining the
   results of many experiments it should be possible to negate their
   impact on the determination of the cosmological parameters.  

   \bigskip
\bigskip 
 
   I wish to thank Mario Livio, Alex Filippenko, and Robert Kirshner for
 helpful discussions and comments.

\vfill \eject
 
\centerline {\bf References}

\refitem Aguirre, A. 1999a, ApJ, 512, 19
 
\refitem Aguirre, A. 1999b, ApJ, 525, 583

\refitem Alard, C. \& Lupton, R. H., 1998, ApJ, 503, 325
 
\refitem Aldering, G., Knop, R., \& Nugent, P. E., 2000, accepted ApJ (astro-ph/0001049)

\refitem Aldering, G., et al., 1998, IAU Circ. 7046

\refitem Aldering, G., et al., 2000, private communication

\refitem Bahcall, J. N.  et al. 1996, ApJ, 457,19
 
\refitem Bahcall, N. A., Fan, X., \& Cen, R., ApJ, 1997, 485, 53

\refitem Balbi, A., et al., 2000, astro-ph/0005124

\refitem Barber, A. J., 2000, astro-ph/0003304

\refitem Branch, D., ARAA, 1998, 36, 17

\refitem Branch, D., Romanishan, W., \& Baron, E., 1996, ApJ, 465, 73

\refitem Burstein, D., \& Heiles, C. 1982, AJ, 87, 1165

\refitem Cappellaro, E., Turatto, M., Tsvetkov, D. Yu., et al., 1997,
A\&A, 322,431

\refitem Carlberg, R. G., Yee, H. K. C., Ellingson, E., Abraham, R., Gravel, P., Morris, S. \& Pritchet, C. J., 1996, ApJ, 462, 32

\refitem Caldwell, R. R., Dave, R., \& Steinhardt, P. J., 1998,
Ap\&SS, 261, 303
 
\refitem Carroll, S. M., Press, W. H., \& Turner, E. L., 1992, ARAA, 30, 499

\refitem Colgate, S. 1979, ApJ, 232, 404

\refitem Colless, M., Saglia, R. P., Burstein, D., Davies, R.,
McMahan, R. K., \& Wegner, G., 1999, in {\it Cosmic Flows Workshop} 
astro-ph/9909062 

\refitem Courteau, S., Willick, J. A., Strauss, M. A., Schlegel, D.,
\& Postman, M., 2000, ApJ, submitted (astro-ph/0002420)

\refitem Dale, D. A., Giovanelli, R., Haynes, M. P., Campusano, L. E.,
\&  Hardy, E., 1999, AJ, 118, 1489 

\refitem de Bernardis, P., et al., 2000, Nature, 404, 955

\refitem Dominguez, I., H\"{o}flich, P., Straniero, O., \& Wheeler, C.,
1999, in {\it Future Directions of Supernovae Research: Progenitors to
Remnants} (Assergi, Italy)

\refitem Drell, P. S., Loredo, T. J., \& Wasserman, I., 2000, ApJ,
530, 593

\refitem Eigenson, M. 1949, Astron. Zh., 26, 278

\refitem Eisenstein, D., Hu, W., \& Tegmark, M., 1999, ApJ, 518, 2 

\refitem Falco, E. et al. 1999, ApJ, 523, 617

\refitem Feldman, H. A., \& Watkins, R., 1998, ApJ, 494, 129

\refitem Filippenko, A. V. et al. 2000, in preparation

\refitem Filippenko, A.V. 1997, ARA\&A, 35, 309

\refitem Frieman, J. A., 1996, astro-ph/9608068

\refitem Garnavich, P., et al. 1998a, ApJ, 493, 53

\refitem Garnavich, P., et al. 1998b, ApJ, 509, 74

\refitem Gilliland, R. L., Nugent, P. E., \& Phillips,  M. M., 1999,
ApJ, 521, 30

\refitem Goldhaber, G. 1998, B.A.A.S., 193, 4713
 
\refitem Goldhaber, G., et al. 1997, in {\it Thermonuclear Supernovae}, eds. P. Ruiz-Lapuente, R. Canal,  \& J. Isern (Dordrecht: Kluwer), p. 777

\refitem Goobar, A. \& Perlmutter, S., 1995, ApJ, 450, 14

\refitem Goodwin, S. P., Thomas, P. A., Barber, A. J., Gribbin, J., \&
Onuora, L. I., 1999, astro-ph/9906187
 
\refitem Groom, D. E. 1998, B.A.A.S., 193, 11102

\refitem Hamuy, M., Phillips, M. M., Maza, J. Suntzeff, N. B., Schommer, 
R.A. \& Avil\'es, R. 1995, AJ, 109, 1

\refitem Hamuy, M., et al. 1996a, AJ, 112, 2408
 
\refitem Hamuy, M., et al. 1996b, AJ, 112, 2438

\refitem Hamuy, M. \& Pinto, P. A., 1999, AJ, 117, 1185

\refitem Hanany, S., et al. 2000, astro-ph/0005123

\refitem Hatano, K., Branch, D., \& Deaton, J., 1998, ApJ, 502, 177
 
\refitem H\"{o}flich, P., Wheeler, J. C., \& Thielemann, F. K. 1998,
ApJ, 495, 617

\refitem Hogg, D., \& Turner, E. L.,1998, PASP, 110, 727

\refitem Hogg, D., 2000, astro-ph/0001419
 
\refitem Holz, D. E., 1998, ApJ, 506, 1

\refitem Holz, D. E. \& Wald, R. M., 1998, Phys.Rev., D58, 063501

\refitem Howell, D. A., Wang, L., \& Wheeler, J. C., 2000, ApJ, 530, 166

\refitem Hoyle, F., Burbidge, G., \& Narlikar, J. V., 2000, in {\it A
Different Approach to Cosmology}, Cambridge University Press
(Cambridge: England)

\refitem  Hudson, M. J., Smith, R. J.,  Lucey, J. R., Schlegel, D. J.,
\& Davies, R. L., 1999, ApJ, 512, 79

\refitem Jha, S., et al. 1999, ApJS, 125, 73

\refitem Kantowski, R., Vaughan, T., \& Branch, D. 1995, ApJ, 447, 35

\refitem Kantowski, R., 1998, ApJ, 507, 483
 
\refitem Kim, A., Goobar, A., \& Perlmutter, S. 1996, PASP, 108, 190

\refitem Kirshner, R. P., \& Kwan, J., 1974, ApJ, 193, 27

\refitem Landolt, A. U. 1992, AJ, 104, 340

\refitem Lange, A. E., et al., 2000, astro-ph/0005004

\refitem Lauer, T., \& Postman, M. 1994, Ap.J. 425, 418

\refitem Leibundgut, B., 2000, A\&A, in press

\refitem Leibundgut, B., 1996, ApJ, 466, 21

\refitem Li, W., Filippenko, A. V., Riess, A. G., Treffers, R. R.,
Hu., J., \& Qiu, Y., 2000, ApJ, submitted 

\refitem Li, W., Filippenko, A. V., \& Riess, A. G., 2000, ApJ, submitted 

\refitem Lin, H. et al. 1996, ApJ, 471, 617

\refitem Lineweaver, C. H., 1998, ApJ, 505, 69

\refitem Livio, M., 2000, in {\it Type Ia Supernovae: Theory and
Cosmology} (Cambridge University Press)

\refitem Malmquist, K. G., Medd. Lund Astron. Obs. Ser. II, 32, 64

\refitem Malmquist, K. G., 1936, Stockholm Obs. Medd., no. 26

\refitem Mathis, J. S., 1990, ARA\&A, 28, 37

\refitem McGaugh, S. \& de Blok, W. J. G., 1998a, ApJ, 499, 66

 \refitem McGaugh, S. \& de Blok, W. J. G., 1998b, ApJ, 499, 41

\refitem McLeod, B. A., et al. 1999, in {\it Gravitational Lensing:
Recent Progress and Future Goals}, ed. Brainerd T. \& Kochanek C. S,
ASP (astro-ph/9910065)

\refitem Melchiorri, A., et al., 1999, astro-ph/9911445

\refitem Metcalf, R. B., 1999, MNRAS, 305, 746

\refitem Metcalf, R. B. \& Silk, J., 1999, ApJ, 519, 1

\refitem Mihalas, D., \& Binney, J. 1981, Galactic Astronomy:
Structure and Kinematics (2d ed.; San Francisco: W. H. Freeman)

\refitem Milgrom, M., 1983, ApJ, 270, 371

\refitem Milgrom, M., 1988, ApJ, 333, 689

\refitem Miller, A. D., et al., 1999, ApJ, 524, 1

\refitem Narlikar, J. V. \& Arp, H. C., 1997, ApJ, 482, 119

\refitem Norgaard-Nielsen, H., et al. 1989, Nature, 339, 523

\refitem Parodi, B. R., Saha, A., Sandage, A., \& Tammann, G. A.,
2000, astro-ph/0004063

\refitem Perlmutter, S. et al., 1995, ApJ, 440, 41
 
\refitem Perlmutter, S. et al., 1997, ApJ, 483, 565

\refitem Perlmutter, S., et al. 1998, Nature, 391, 51
 
\refitem Perlmutter, S., et al. 1999, ApJ, 517, 565

\refitem Phillips, M. M. 1993, ApJ, L105, 413
 
\refitem Phillips, M. M., Lira, P.,
 Suntzeff, N. B., Schommer, R. A.,
 Hamuy, M., Maza, J., 1999, AJ, 118, 1766

\refitem Rana, N. C., 1980, Ap\&SS, 71, 123

\refitem Rana, N. C., 1979, Ap\&SS, 66, 173

\refitem Riess, A. G., Press W.H., Kirshner, R.P., 1995, ApJ, 438 L17

\refitem Riess, A. G., Press, W.H., \& Kirshner,  R.P. 1996, ApJ, 473,
88 
 
\refitem Riess, A. G., et al. 1997, AJ, 114, 722
 
\refitem Riess, A. G., et al. 1998, AJ, 116, 1009
 
\refitem Riess, A. G., et al. 1999b, AJ, 118, 2675
 
\refitem Riess, A. G., et al. 1999a, AJ, 117, 707
 
\refitem Riess, A. G., Filippenko, A. V., Li, W., \& Schmidt,
B. P. 1999c, AJ, 118, 2668 

\refitem Riess, A. G., 1999, in {\it Cosmic Flows Workshop, Victoria,
Canada, July 1999}, ed. S. Courteau, M. Strauss \& J. Willick, ASP, astro-ph/9908237

\refitem Riess, A. G., et al. 2000, ApJ, accepted (astro-ph/0001384)

\refitem Roos, M., \& Harun-or-Rashid, S. M., 2000, astro-ph/0003040

\refitem Ruiz-Lapuente, P., \& Canal, R., 1998, ApJ, 497, 57

\refitem Saha, A., et al. 1999, ApJ, 522, 802
 
\refitem Sandage, A., \& Hardy, E., 1973, ApJ, 183, 743

\refitem Schlegel, D. J., Finkbeiner, D. P., \& Davis, M. 1998, ApJ, 500, 525

\refitem Schmidt, B. P., et al. 1998, ApJ, 507, 46

\refitem Schmidt, B. P., 2000, private communication

\refitem Seljak, U., \& Holz, D., 1999, A\&A, 351, 10

\refitem Shaw, R. L., 1979, A\&A, 76, 188

\refitem Shi, X. \& Turner, M. S., 1998, 493, 519 

\refitem Strauss, M.A., \& Willick, J.A., 1995, PhR, 1995, 261, 271

\refitem Tammann, G. A. 1979, in ESA/ESO Workshop on Astronomical Uses
of the Space Telescope, ed. F. Macchetto, F. Pacini, and M. Tarenghi
(Geneva:ESO), 329

\refitem Tegmark, M., \& Zaldarriaga, M., 2000, astro-ph/0002091

\refitem Tonry, J., et al., 1999, IAU Circ. 7312

\refitem Tomita, K., 2000, astro-ph/ 0005031

\refitem Totani, T., \& Kobahashi, C., ApJ, 526, 65 

\refitem Tripp, R. 1997, A\&A, 325, 871

\refitem Tripp, R., \& Branch, D., 1999, ApJ, 525, 209

\refitem Trumpler, R., 1930, PASP, 42, 214

\refitem Turner, M. S., \& Tyson, J. A., 1999, Rev. Mod. Phys., 71, 145 

\refitem Turner, M. S., 1999, PASP, 111, 264 

\refitem Umeda, H. et al. 1999, ApJ, 522, 43

\refitem van den Bergh, S., 1994, ApJS, 92, 219

\refitem von Hippel, T., Bothun, G. D., \& Schommer, R. A., 1997, AJ, 114, 1154
 
\refitem Wagoner, R. V., 1977, ApJ, 214, L5

\refitem Wambsganss, J., Cen, R., Guohong, X., \& Ostriker, J. 1997, ApJ, 475, L81

\refitem Wang, Y., 2000, accepted to ApJ, astro-ph/9907405 

\refitem Wang, Y.,  Spergel, D. N., \& Turner, E. L., 1998, 498, 1

\refitem Watkins, R. \& Feldman, H. A., 1995, ApJ, 453, 73

\refitem Whitmore, B., Heyer, I., \& Casertano, S., 1999, PASP, 111,
1559

\refitem Willick, J. A., 2000, in {\it Proceedings of the XXXVth
Rencontres de Moriond: Energy Densities in the Universe} astro-ph/0003232

\pagestyle{empty}

{\bf Figure Captions:}

Figure 1: Hubble diagrams of SNe Ia from Perlmutter et al. (1999; SCP) and Riess et al. (1998; HZT). Overplotted are three cosmologies: ``low'' and
``high'' $\Omega_M$ with $\Omega_\Lambda=0$ and the best fit for a flat
cosmology, $\Omega_M=0.3$, $\Omega_\Lambda=0.7$.  The bottom panel
shows the difference between data and models from the $\Omega_M=0.3$,
$\Omega_\Lambda=0$ prediction. The
average difference between the data and the $\Omega_M=0.3$,
$\Omega_\Lambda=0$ prediction is 0.28 mag.

Figure 2: SNe Ia joint confidence intervals for ($\Omega_M$,$\Omega_\Lambda$) from Perlmutter et al. (1999; SCP) and Riess et al. (1998; HZT).  Regions
representing specific cosmological scenarios are indicated.

Figure 3: Type Ia supernova model spectral energy distributions for solar and ${1 \over 3}$ metallicities.  Superimposed are the transmission functions for standard passbands; from left to right is $U,B,V$, and $R$.  From H\"{o}flich, Wheeler \& Thielemann (1998)

Figure 4: The nearby Hubble diagram of SNe Ia in different host galaxy morphologies.  In the top panel the SNe Ia are treated as standard candles and in the bottom panel distances are determined with the MLCS method (Riess, Press, \& Kirshner 1996; Riess et al. 1998).  After MLCS corrections are made, the distance Hubble flow residuals are independent of host galaxy morphology.

Figure 5: Spectra of 4 nearby and 1 high-redshift SN Ia at the same phase (Riess et al. 1998).   Within the observed variance
of SN Ia spectral features, the spectra of high-redshift SNe Ia are
 indistinguishable from the low-redshift SNe Ia.  The spectra of the low-redshift SNe Ia were resampled and convolved with Gaussian noise to match
the quality of the spectrum of SN 1998ai.

Figure 6: Comparison of color excess, $E_{B-V}$, for nearby SNe Ia from the Calan/Tololo survey (Hamuy et al. 1996a,b) and high-redshift SNe Ia from Perlmutter et al. (SCP; 1999).  The color excesses at high and low redshift are consistent indicating negligible extinction and/or no evidence for color evolution.

Figure 7: The color evolution, $B-I$, and the color excess, $E_{B-I}$, of a h
igh-redshift SN Ia, SN
1999Q ($Z=0.46$), compared to the custom MLCS template curve with no dust and enough dust (of either
Galactic-type or greyer) to nullify the cosmological constant.  
 The smaller error bars are from photometry noise; the larger error
 bars include all sources of uncertainty such as intrinsic dispersion of SN Ia $B-I$ color, $K$-corrections, and photometry zeropoints.  The
data for SN 1999Q are consistent with no reddening by dust, moderately
inconsistent with $A_V$=0.3 mag of grey dust (i.e., graphite dust with
minimum size $> 0.1 \ \mu$m; Aguirre 1999a,b) and $A_V$=0.3 mag of
Galactic-type dust.  From Riess et al. (2000).

Figure 8: Probability distribution $P(\mu)$ for supernova apparent brightness $\mu$ normalized to $\mu=1$ for a filled beam (i.e., a homogeneous Universe).  The vertical lines are at the empty-beam value.  ``Galaxies'' are treated as isothermal spheres and truncated at a radius of 380 kpc; ``compact objects'' are point masses.  From Holz (1998.)

Figure 9: Cosmological constraints from Perlmutter et al. (SCP; 1999) for three weak lensing scenarios.  Fit C assumes a filled beam, case K assumes an empty beam, and fit L is a model with weak lensing by up to 
$\Omega_M=0.25$ in compact objects.

Figure 10: The projected
   distances from the host centers of nearby SNe Ia discovered photographically and with CCDs and high-redshift SNe Ia discovered with CCDs (Howell, Wang, \& Wheeler 2000).

Figure 11: MLCS distance residuals from the Hubble flow for nearby SNe Ia versus their projected distance from their host centers (Riess et al. 1999a).

Figure 12: The Hubble diagram of SNe Ia (see Figure 1) and the effect of a systematic
 error which grows linearly with redshift (e.g., evolution or grey extinction).

Figure 13: Cosmological constraints from SNe Ia, CMB, and matter (Turner 1999).

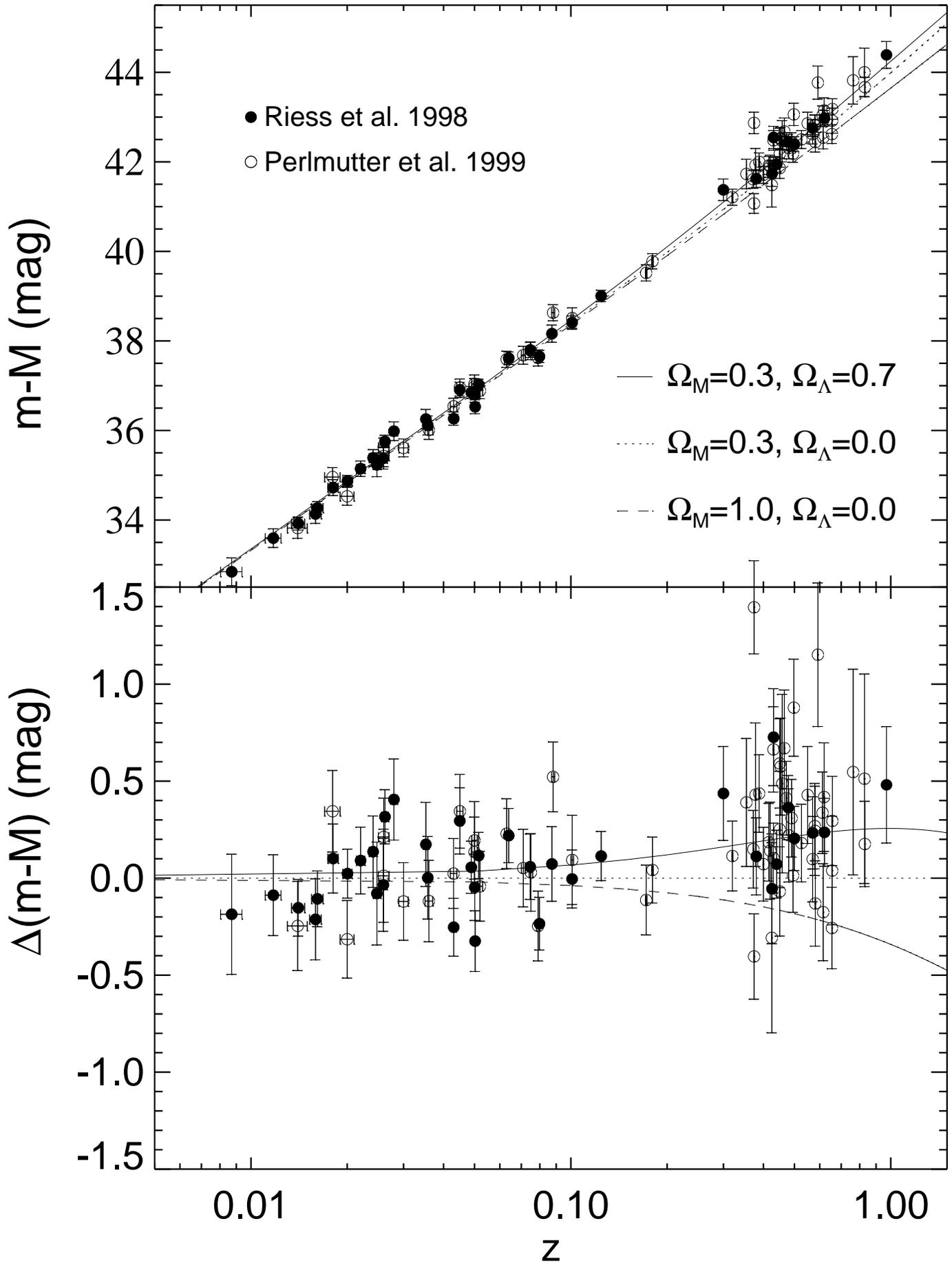
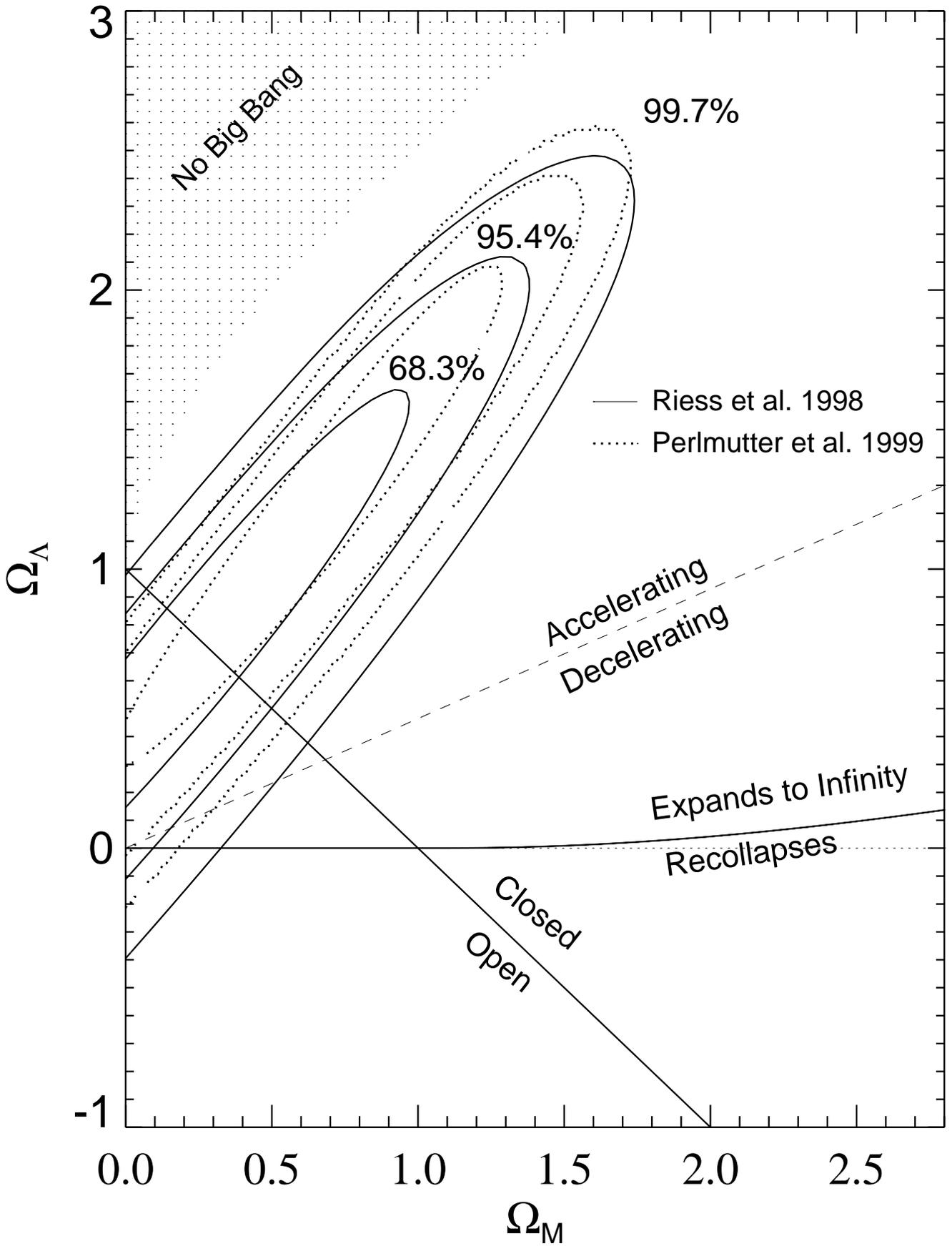
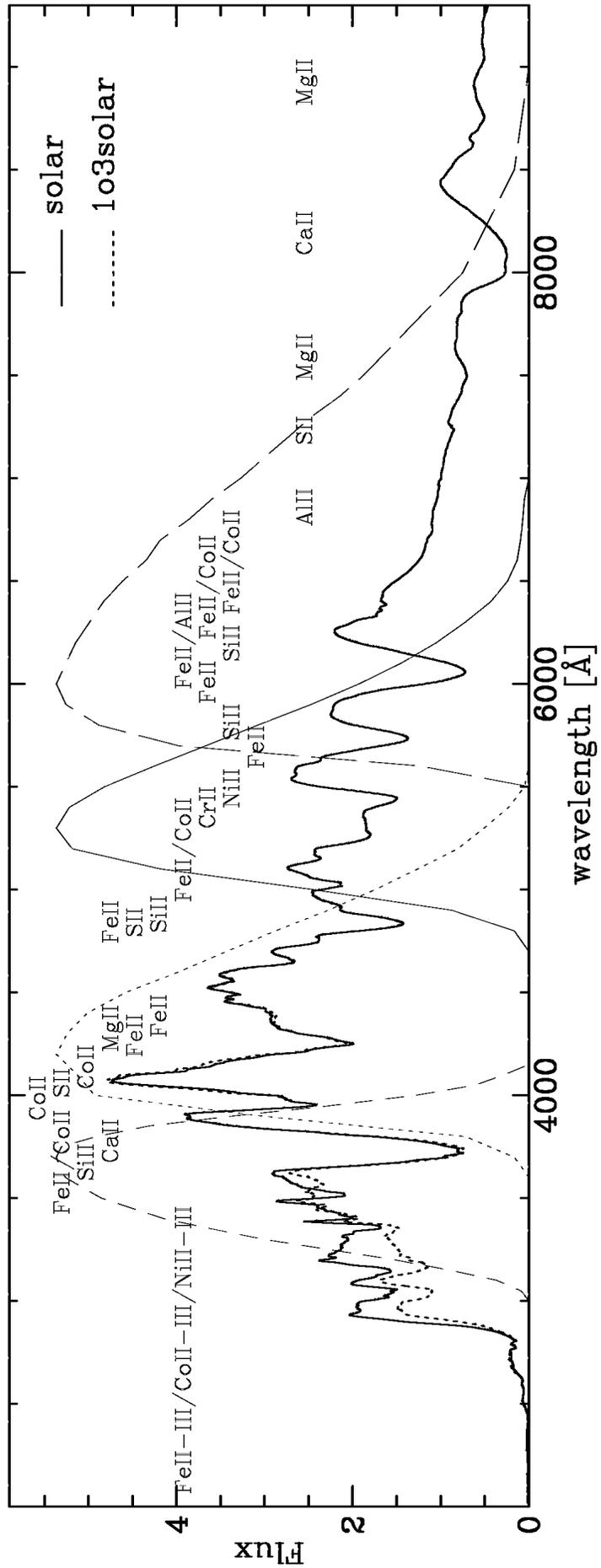
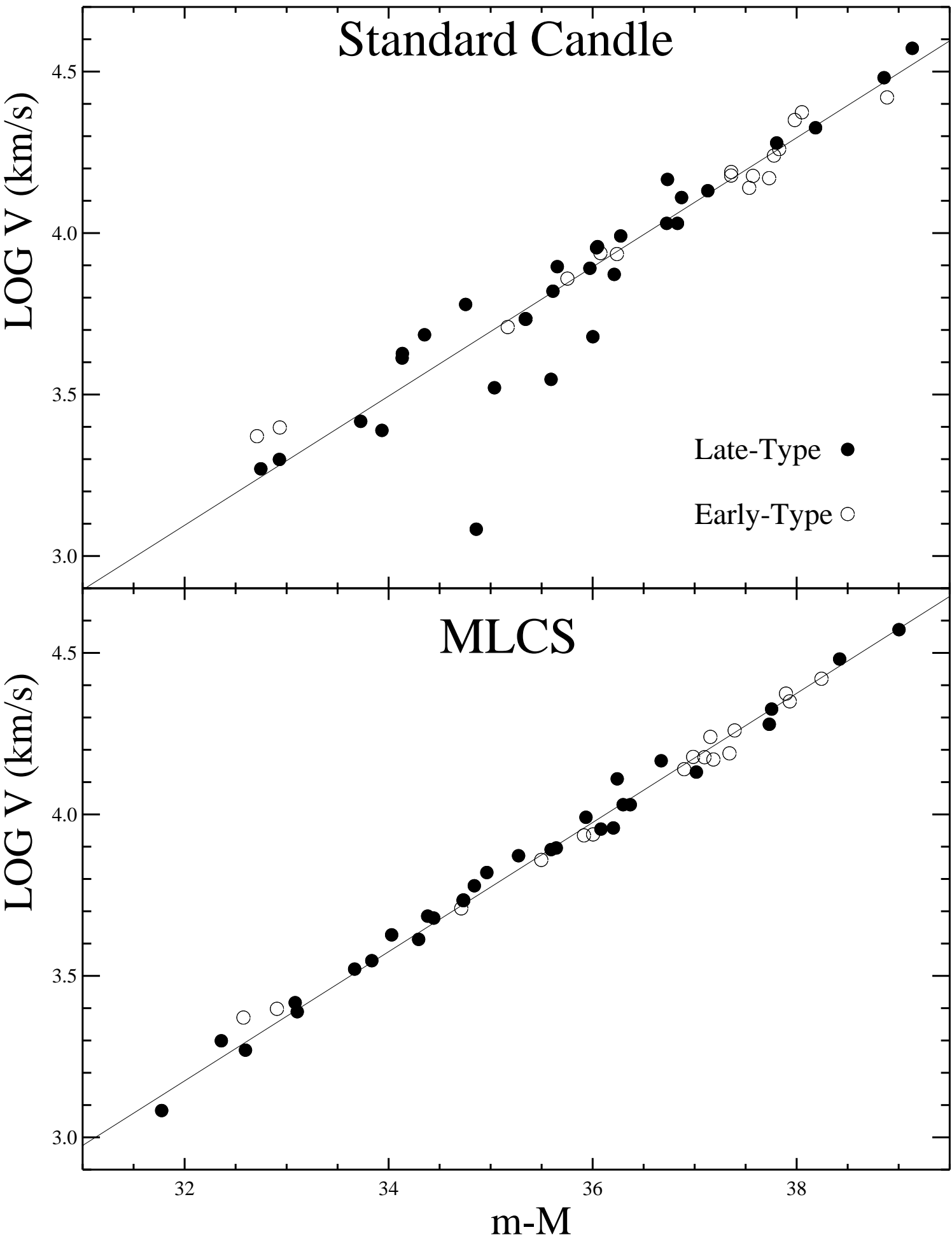
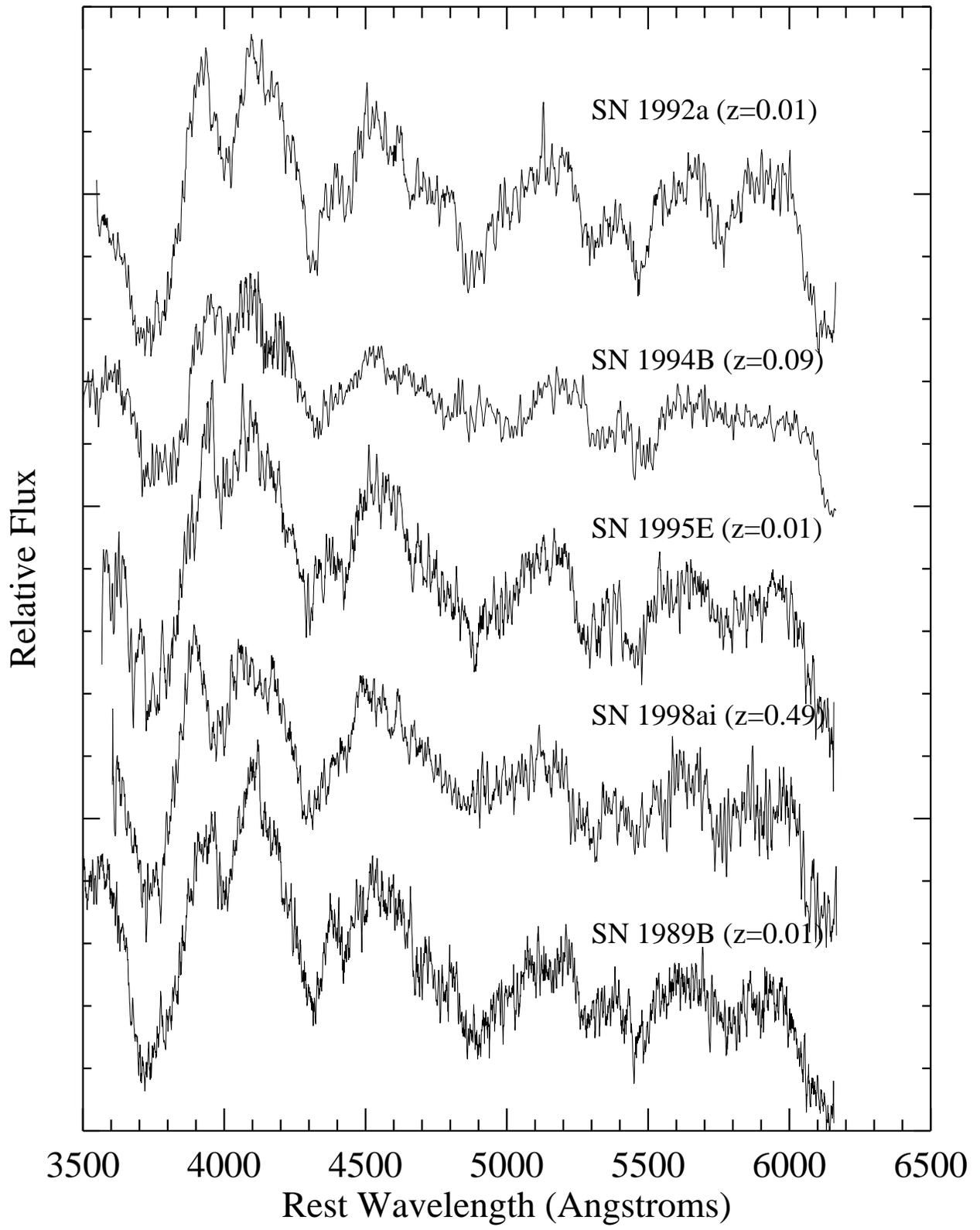
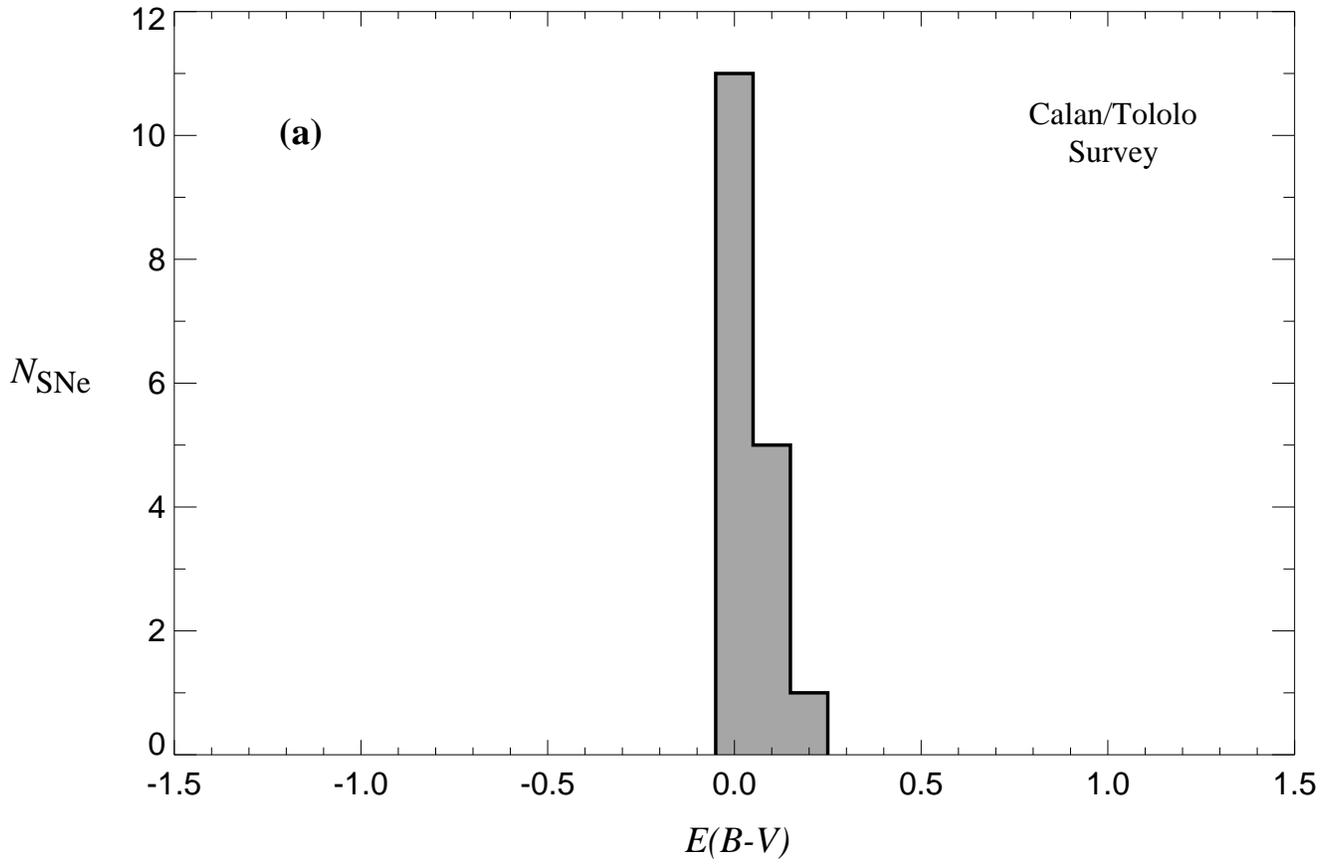
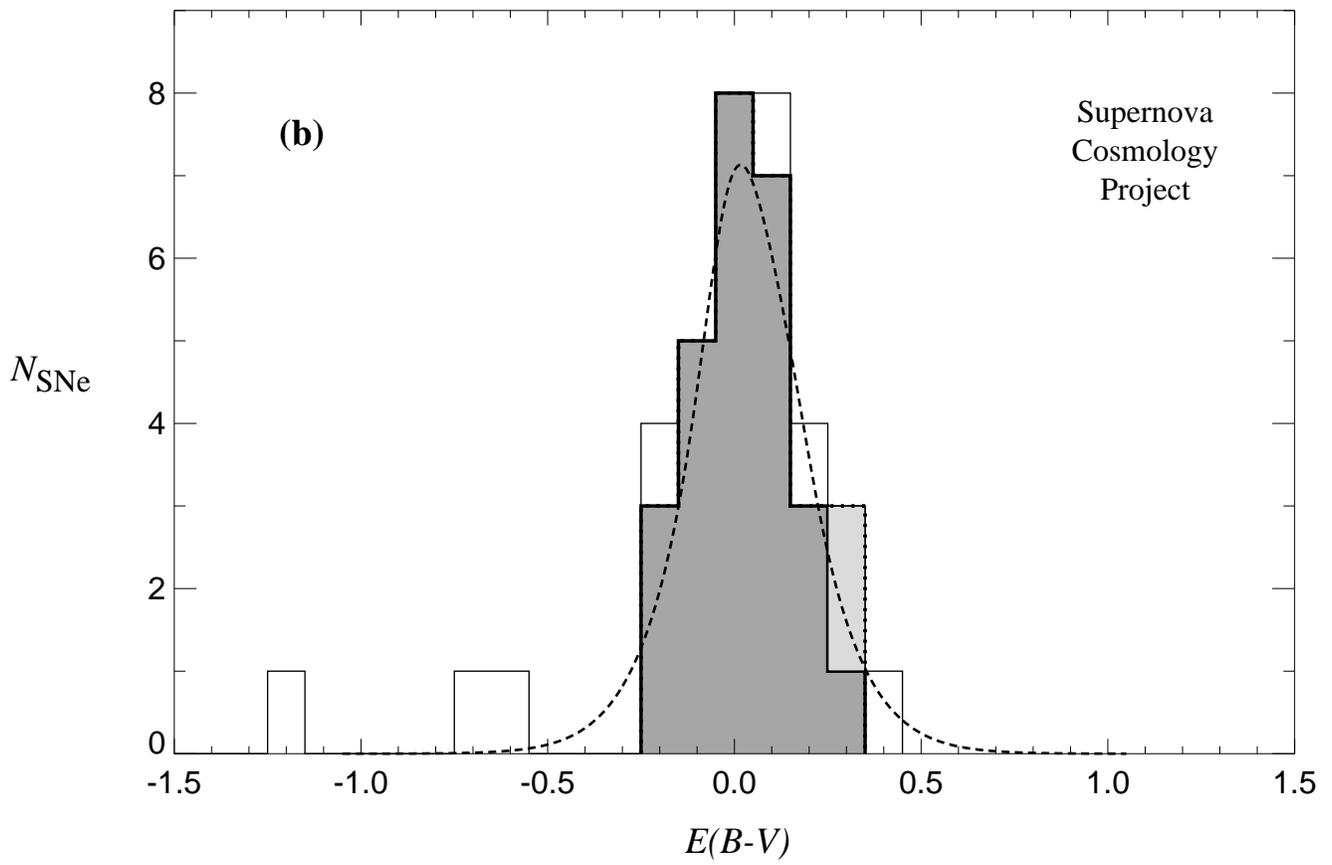
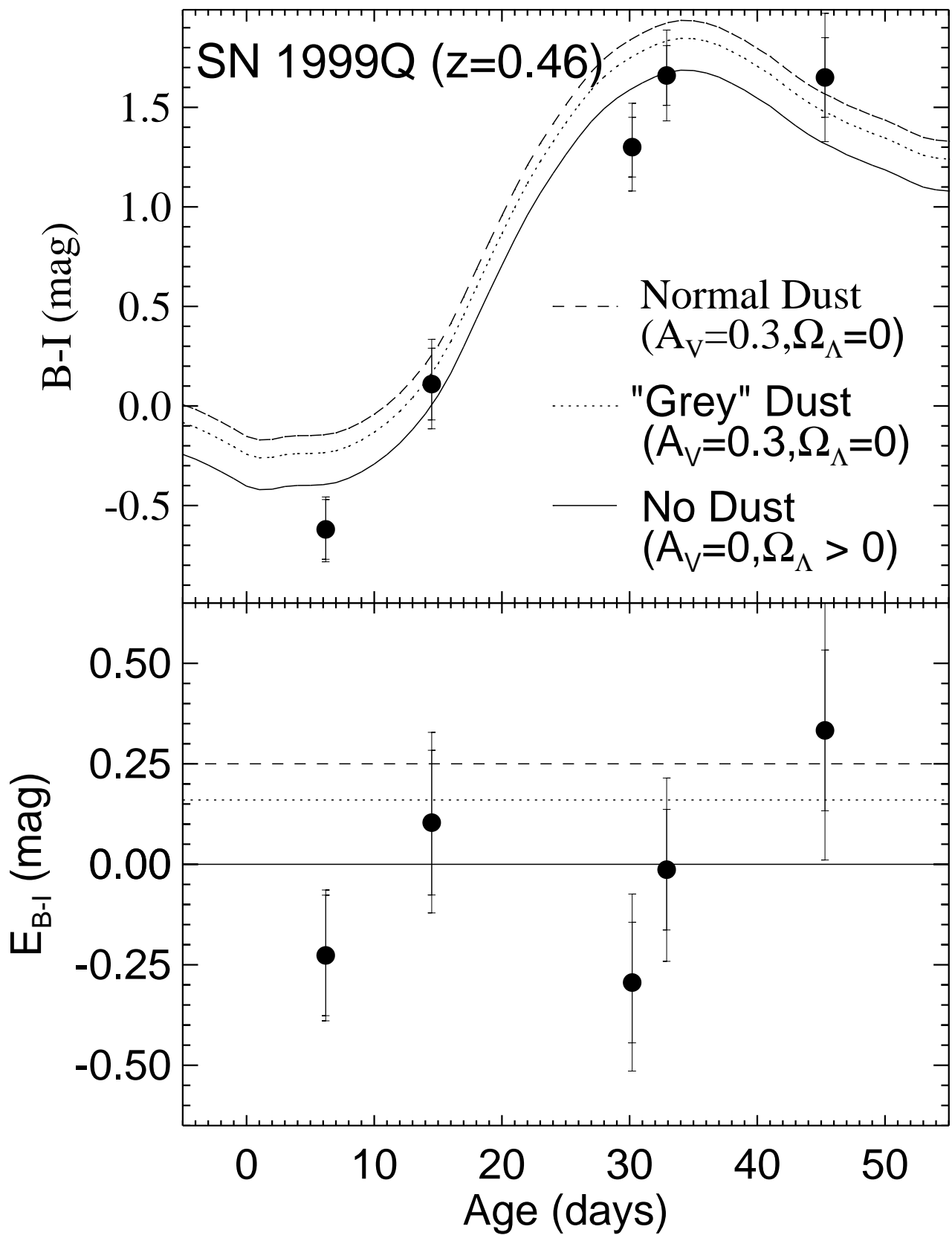
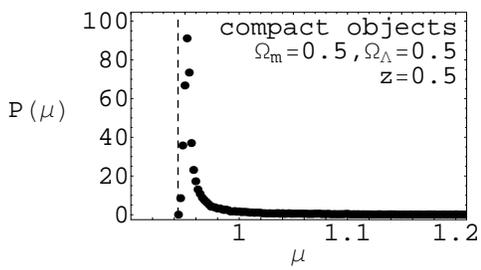
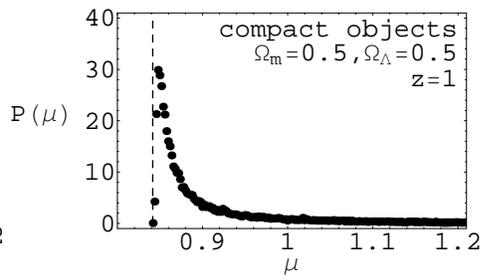
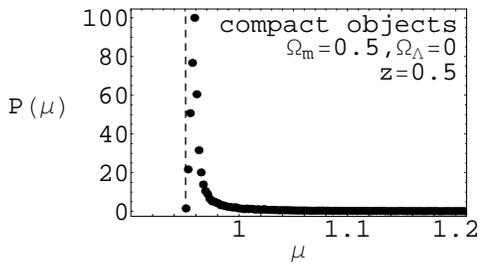
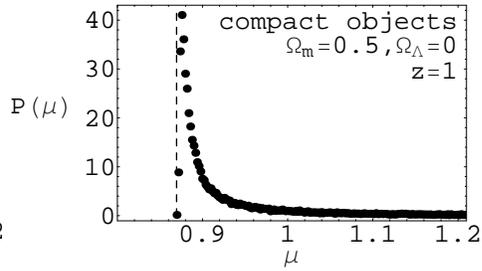
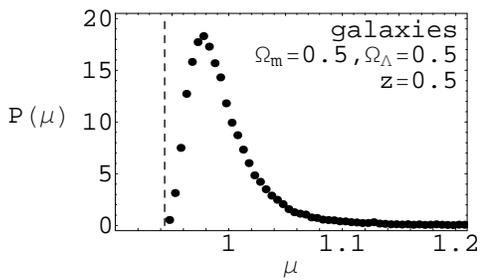
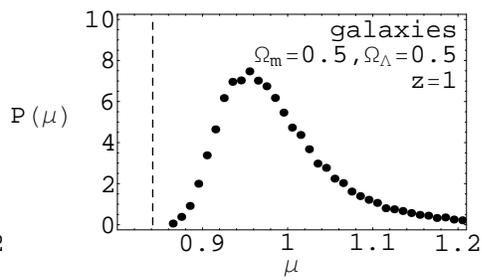
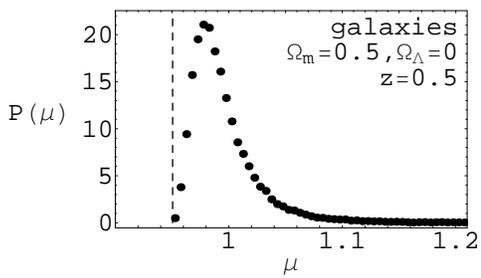
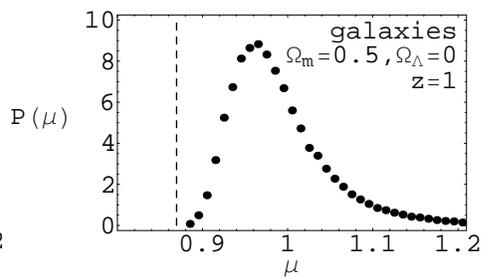
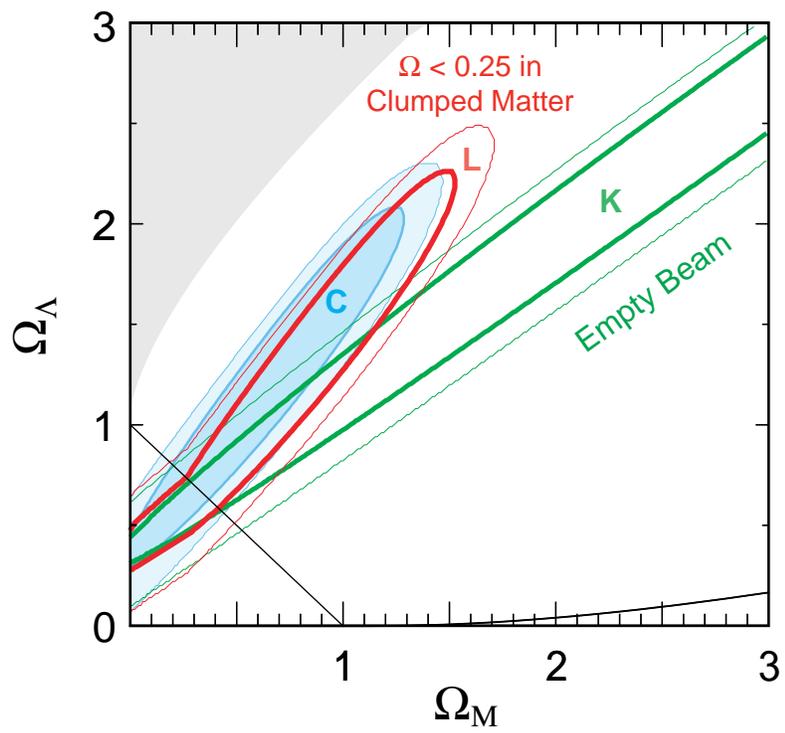
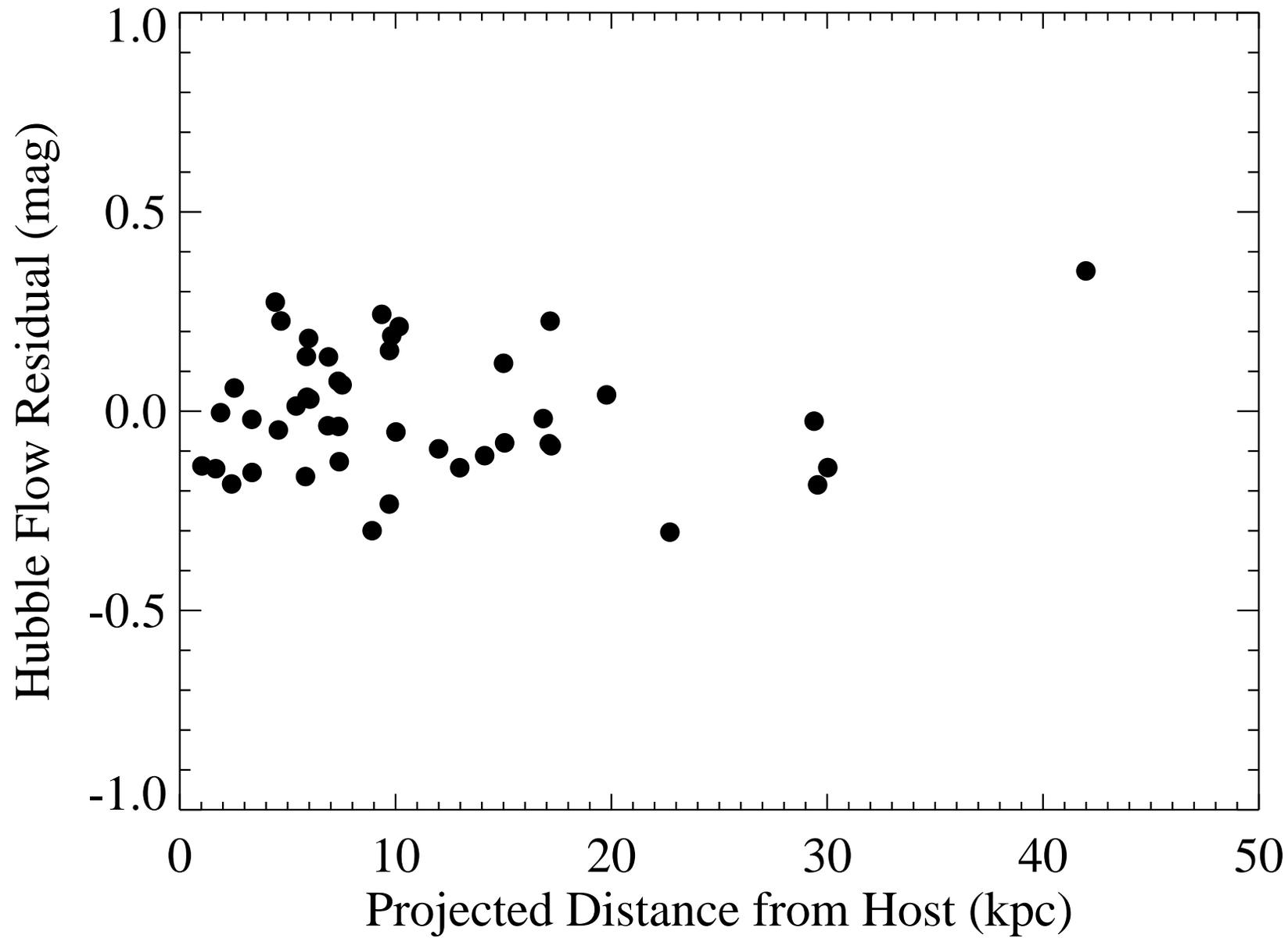
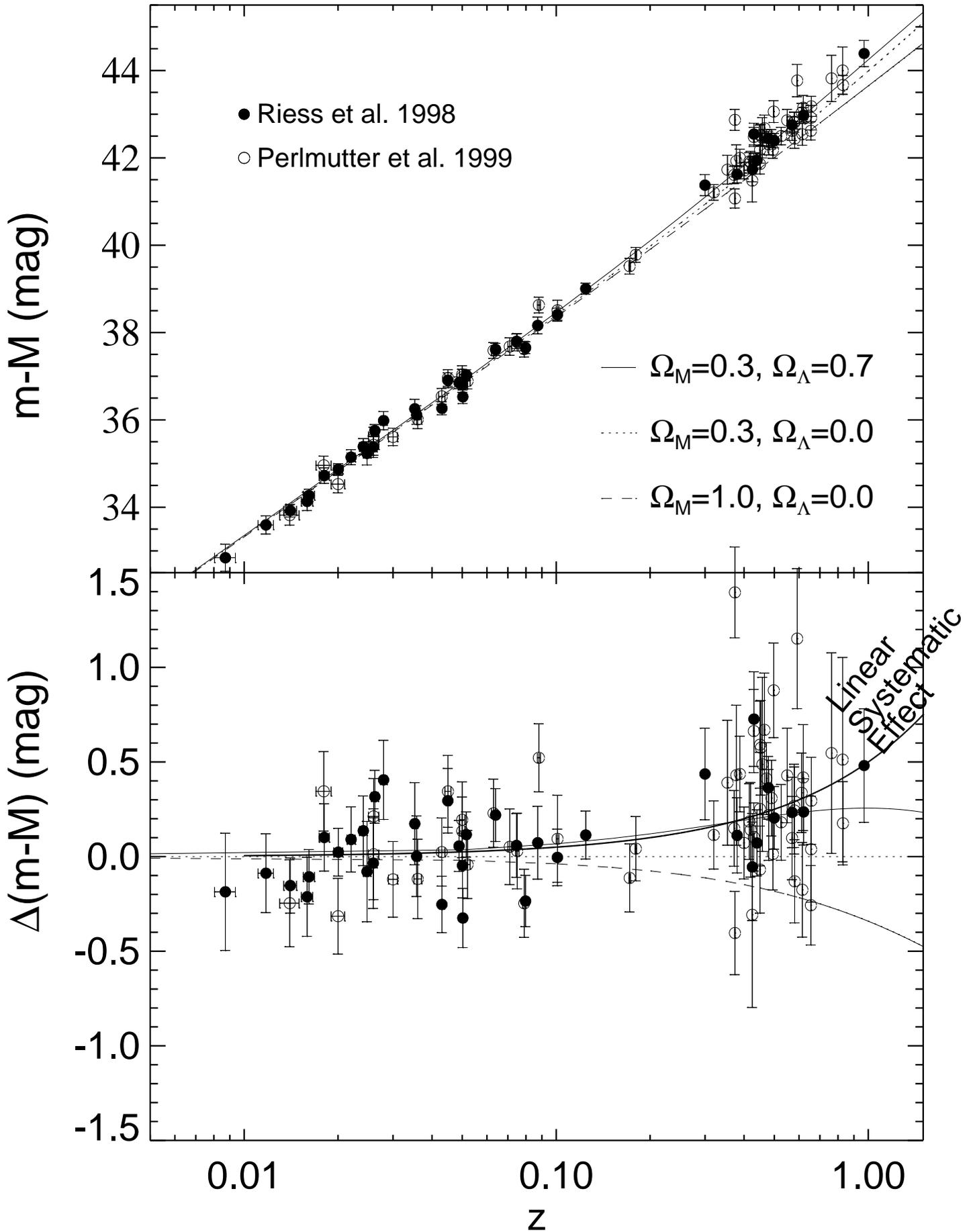
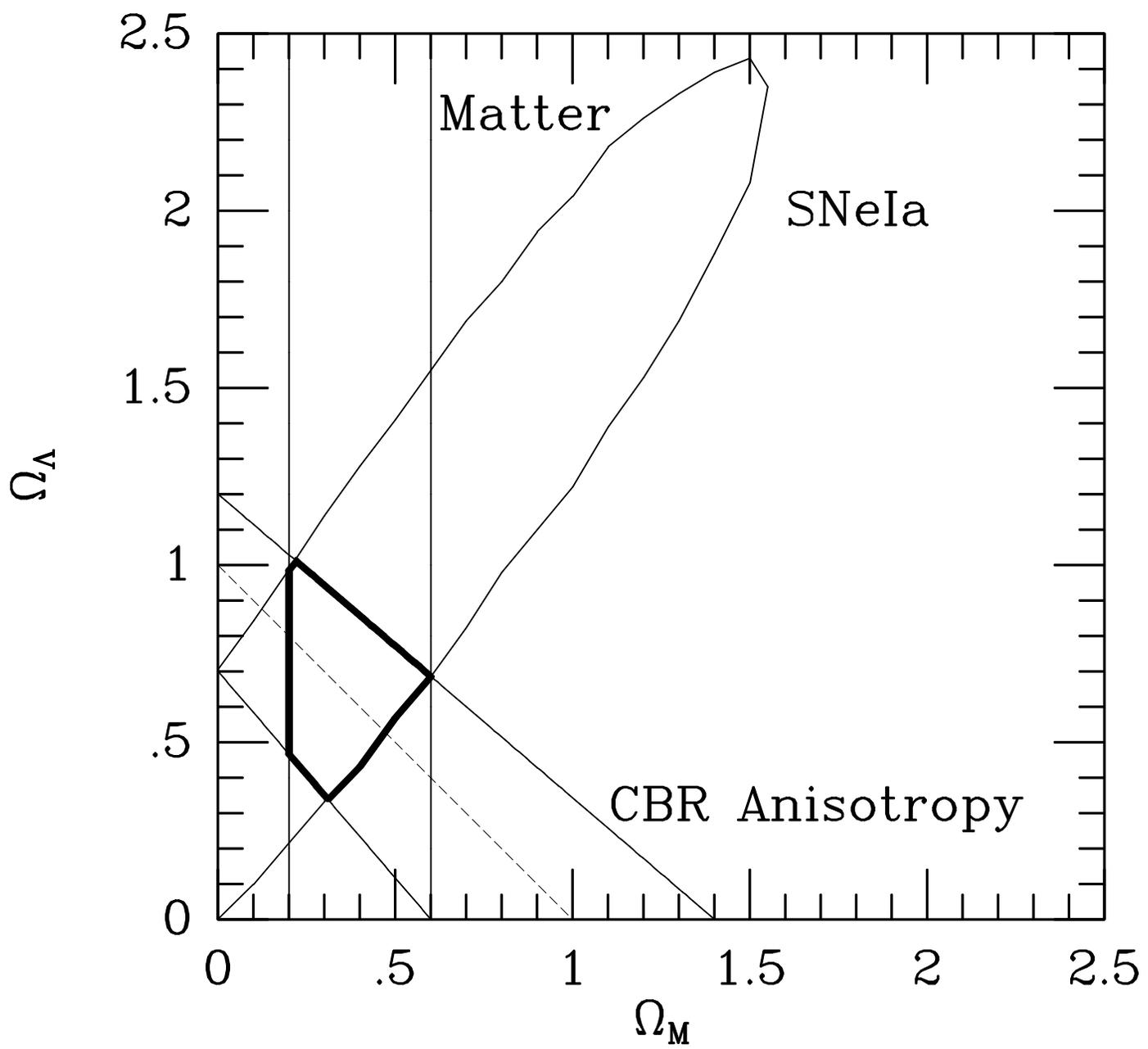
\end{document}